\documentclass[preprint,amsmath,amssymb,aps]{revtex4-1}
\usepackage[english]{babel}
\usepackage{txfonts}
\usepackage[latin1]{inputenc} %% Schriftsatz
\usepackage{textcomp}     %% Text Companion Paket
\usepackage[T1]{fontenc}
\usepackage{comment}
\usepackage{tikz}

\newcommand\scalemath[2]{\scalebox{#1}{\mbox{\ensuremath{\displaystyle #2}}}} %allows to down-scale matrices
\usepackage{natbib}

\usepackage{hyperref}
\usepackage{xcolor}
\usepackage{cleveref}                               % allows to reference multiple equations
\definecolor{dark-red}{rgb}{0.4,0.15,0.15}
\definecolor{dark-blue}{rgb}{0.15,0.15,0.4}
\definecolor{medium-blue}{rgb}{0,0,0.5}
\hypersetup{
colorlinks, linkcolor={dark-blue},
citecolor={dark-blue}, urlcolor={medium-blue}
}
\usepackage[abs]{overpic}

\newcommand{\eq}[1]{Eq.\,\eqref{#1}}

\renewcommand{\l}{\left(}
\renewcommand{\r}{\right)}

\renewcommand{\b}[1]{\boldsymbol{#1}}

\usepackage{graphicx}
\graphicspath{{pics/}} % Dort liegen die Bilder des Dokuments 
\usepackage{wrapfig}
\usepackage[caption=false]{subfig}
\usepackage{subfloat}
\usepackage{isotope}
\usepackage{dcolumn}
\usepackage{color}
\usepackage{hyperref} %% Verweise
\usepackage{placeins}
\DeclareGraphicsExtensions{.pdf,.png,.jpg}

\usepackage{empheq} %mehrere aligned equations mit boxen

\usepackage{mathrsfs} % Zusatzzeichen
\usepackage{amssymb}
\usepackage{amsfonts}
\usepackage{bbm}
\usepackage{bbold}
\usepackage{amsmath}
\usepackage{empheq} 
\usepackage{braket}
\usepackage{mathtools}
\makeatletter
\newcommand{\subalign}[1]{%
  \vcenter{%
    \Let@ \restore@math@cr \default@tag
    \baselineskip\fontdimen10 \scriptfont\tw@
    \advance\baselineskip\fontdimen12 \scriptfont\tw@
    \lineskip\thr@@\fontdimen8 \scriptfont\thr@@
    \lineskiplimit\lineskip
    \ialign{\hfil$\m@th\scriptstyle##$&$\m@th\scriptstyle{}##$\crcr
      #1\crcr
    }%
  }
}
\makeatother

\begin{document}

\title{\mbox{Quantum-optical magnets with competing short- and long-range interactions:} \\Rydberg-dressed 
spin lattice in an optical cavity}

\author{Jan Gelhausen}
\email{jg@thp.uni-koeln.de}

\author{Michael Buchhold}

\author{Achim Rosch}

\author{Philipp Strack}

\email{strack@thp.uni-koeln.de}
\homepage{http://www.thp.uni-koeln.de/~strack/}

\affiliation{Institut f\"ur Theoretische Physik, Universit\"at zu K\"oln, D-50937 Cologne, Germany}
%\affiliation{$^{1}$Institut f\"ur Theoretische Physik, Universit\"at zu K\"oln, D-50937 Cologne, Germany}
%\affiliation{$^{2}$Department of Physics, Harvard University, Cambridge, MA 02138, USA}

\date{\today}

\begin{abstract}
\noindent The fields of quantum simulation with cold atoms \cite{bloch12}
and quantum optics \cite{gardiner15} are currently being merged.
In a set of recent pathbreaking experiments with
atoms in optical cavities \cite{landig16,klinder15}, lattice quantum many-body systems 
with both, a short-range interaction 
and a strong interaction potential of infinite range --mediated by a quantized 
optical light field-- were realized. A theoretical modelling of these systems 
faces considerable complexity at the interface of: 
(i) spontaneous symmetry-breaking and emergent phases of 
interacting many-body systems with a large number of atoms $N\rightarrow \infty$, 
(ii) quantum optics and the dynamics of 
fluctuating light fields, and (iii) non-equilibrium physics of 
driven, open quantum systems.
Here we propose what is possibly the simplest, 
quantum-optical magnet with competing short- and long-range interactions, in which all three 
elements can be analyzed comprehensively: a Rydberg-dressed spin lattice \cite{zeiher16} 
coherently coupled to a single photon mode. Solving a set of coupled even-odd sublattice 
master equations for atomic spin and photon mean-field amplitudes, we find three key results. (R1): 
Superradiance and a coherent photon field appears in combination with spontaneously broken 
magnetic translation symmetry. The latter is induced by the short-range 
nearest-neighbor interaction from weakly admixed Rydberg levels. (R2): 
This broken even-odd sublattice symmetry leaves its imprint in the light via a novel peak in the cavity spectrum 
beyond the conventional polariton modes.
(R3): The combined effect of atomic spontaneous emission, drive, and interactions can lead 
to phases with anomalous photon number oscillations.
Extensions of our work include nano-photonic crystals coupled 
to interacting atoms and multi-mode photon dynamics in Rydberg 
systems.

\end{abstract}

\maketitle
{\tableofcontents}

\section{Introduction and key results}
The fields of quantum simulation with cold atoms \cite{bloch12}
and quantum optics \cite{gardiner15} are currently being merged.
On the one hand, to coherently couple photons to low-entropy, correlated 
quantum many-body states --the objects of desire of quantum simulators--
offers new possibilities to imprint atomic coherences and 
quantum correlations onto quantum light such as dissipatively 
anti-bunched photons \cite{peyronel12} or a ``many-fermion'' EIT 
\footnote{Electromagnetically Induced Transparency} 
window without light absorption \cite{piazza14a}.
On the other hand, the inter-atomic interactions mediated by the photons 
opens up explorations of previously inaccessible phases such as
long-ranged quantum spin- and charge glasses \cite{strack11,sarang11,
habibian13,bunti13, mueller12}, bond-ordered phases \cite{benitez16}, dynamical spin-orbit couplings 
\cite{deng14,pan15}, or topological states carrying perpetual currents 
with dynamical gauge couplings \cite{kollath16,zheng16}.

A major objective in the field of quantum simulation is to prepare and probe 
low-entropy quantum magnets. Existing efforts have
focussed on magnetic interactions via 
superexchange \cite{trotzky08}, mapping to charge degrees of freedom
\cite{simon11}, dipolar interactions \cite{lahaye09} with polar molecules \cite{yan13},
magnetic atoms \cite{baier16}, and laser-dressed Rydberg atoms \cite{zeiher16}. 
More complex interaction potentials necessary for 
frustrated magnetism
have also been proposed \cite{bijnen14,glaetzle14,glaetzle15}; optical 
pumping schemes should allow to access non-equilibrium magnets, 
too \cite{tony11,tony13}. One common goal of these efforts is the 
engineering of a \emph{single magnetic interaction rate with a certain angle-dependence and range}, 
which can compete with kinetic energies, longitudinal fields, and the decay processes.

In this article, we want to initiate the study of \emph{quantum-optical magnets with competing 
short- and long-range interactions}, the latter being mediated by a dynamical photon field.
One may wonder how a single spin or quantum dipole (with in principle fixed charge distribution) 
can interact via \emph{two competing potentials with drastically different range} 
and independently tunable magnitude: At the core of our proposal is an atomic 
"two dipoles in-one" unit (illustrated below in Fig.~\ref{Fig:RydbergDressingScheme}), 
to which --depending on the principal quantum number and electronic 
transitions used-- two different force-mediating photon fields can couple simultaneously 
(described in Sec.~\ref{sec:model}).

In addition to the novel magnetic phases an array of such "two dipoles in-one" 
can attain, a key question is how the quantum dynamics of the photon field 
is affected by the magnetic correlations. 
Using an optical cavity for one 
of the photonic force carriers and a deep optical lattice
to freeze out the motion of the atoms is a natural 
experimental set-up, which will allow non-destructive detection of purely magnetic 
correlations via the cavity output spectrum \cite{landig16,klinder15,neuzner16}.
Let us note that the question of how quantum light interacts with a self-interacting set of 
qubits is of broader relevance including for example cavity Rydberg polaritons 
\cite{ningyuan16,pupillo16}, Rydberg-EIT setups \cite{peyronel12,maxwell13,tresp15} 
and nano-photonic devices\cite{rauschenbeutel07,thompson12,rauschenbeutel14}. 
Especially in reduced dimensions with confined 
electric fields, even small qubit-qubit interactions can have a huge effect. 

These systems generate a lot of complexity at the interface of 
three typically only weakly connected areas of physics:
(i) \emph{emergent phases and spontaneous symmetry-breaking} of 
interacting many-body systems in the thermodynamic limit 
($N\rightarrow \infty$ number of qubits)
(ii) \emph{quantum optics and the dynamics of 
fluctuating light fields} (from $M=1$ to $M=\infty$ photon modes), and 
(iii) \emph{non-equilibrium physics of 
driven, open quantum systems}, due to drive and multiple loss channels 
such as photon decay with rate $\kappa$ and atomic spontaneous emission 
with rate $\gamma$.
%
%In particular, (iii) implies that typically one can neither base the theoretical description 
%on $T=0$ ground state, nor on finite-$T$ thermal states. 

The goal of the present paper is to provide a ``base case'' or the 
simplest prototype of a quantum-optical magnet with 
competing short- and long-range interactions, in which the interplay of 
the above mentioned (i)-(iii) can be transparently studied. 

\subsection{Model: Rydberg-dressed spin lattice coupled to single-mode optical light field}

\begin{figure}
\begin{tikzpicture}
\node[inner sep=0pt]
    {\includegraphics[width=12cm]{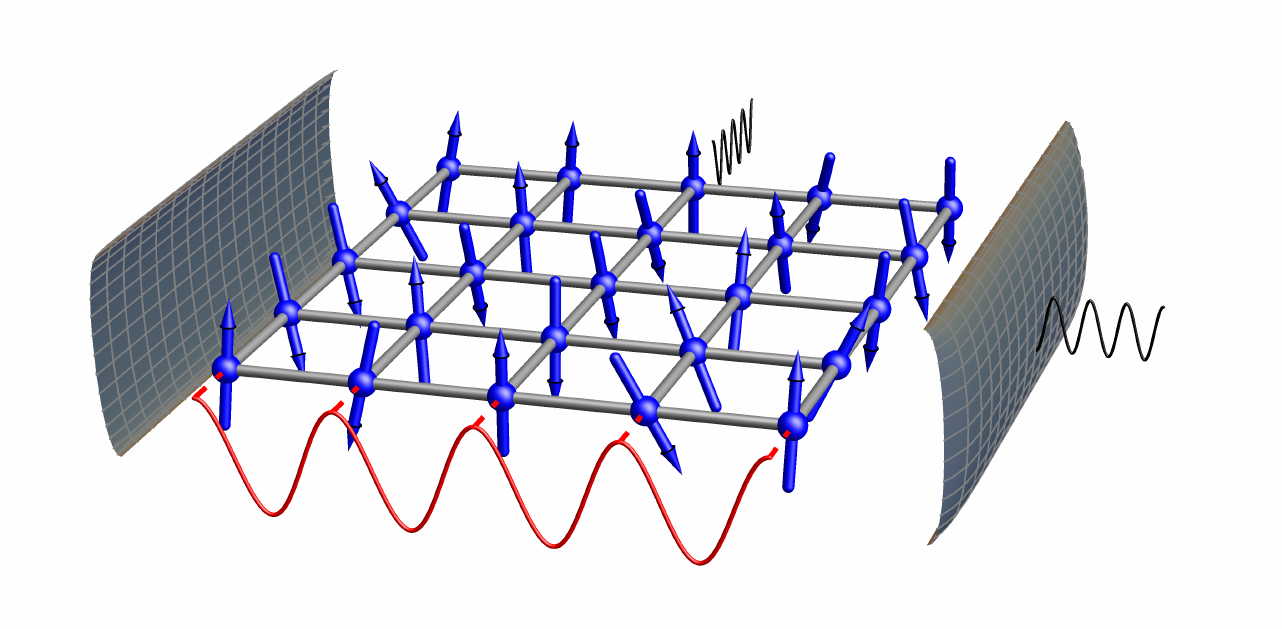}};
    \node[] at (-1.3,-3)
    {\includegraphics[width=4cm,keepaspectratio]{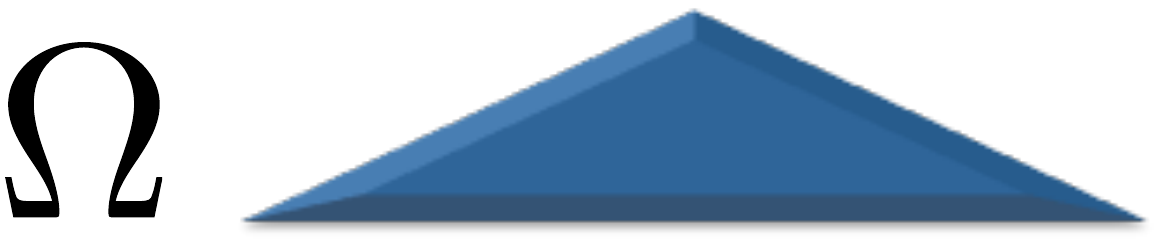}};
%    \node[] at (-6.,1)
%    {\includegraphics[width=2cm,keepaspectratio]{DecayRate.pdf}};
        \node[] at (-1.2,2.2)
         {\includegraphics[width=1.3cm,keepaspectratio]{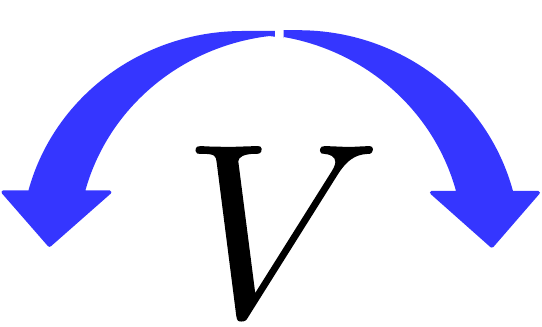}};
        \node[] at (2.2,-1.5)
    {\includegraphics[width=1.3cm,keepaspectratio]{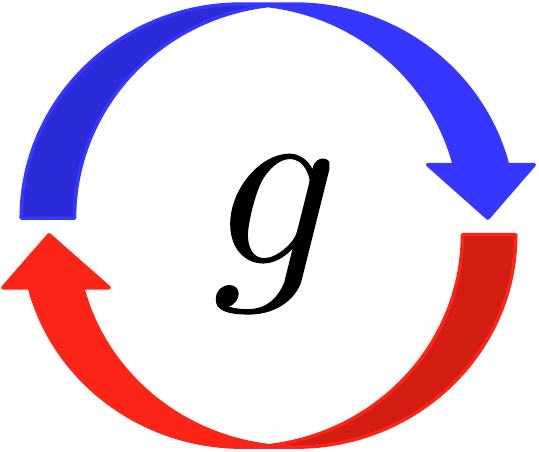}};
%    \begin{scope}[scale=2,transform shape]
%\node at (2.2,-.35) {$\kappa$};
%\end{scope}
    \node[font=\fontsize{22}{24}\selectfont] at (4.4,-.8) {$\kappa$};
        \node[font=\fontsize{22}{24}\selectfont] at (1.5,2.) {$\gamma$};
%\node[] at (2,-2) {\textbf{g}};
%\node[font=\fontsize{25}{48}\selectfont] at (-2,-3.1) {$\Omega$};
%\draw [<->] (2.6,-2) arc (180:10:10pt); 
%\draw[thick]
%    (0,0) node{} 
% -- (1,1) node{} 
% -- (0,2) node{} 
% -- cycle;
\end{tikzpicture}
\caption{Rydberg-dressed spin lattice coupled to single-mode optical light field. 
We take the atoms to be in a deep optical lattice (for example a Mott insulator at 
unit filling) such that no motion occurs and only internal spin excitations drive 
the dynamics. From weakly admixing a Rydberg level, an effective 
nearest-neighbor interaction (repulsive) $V$ competes with an effectively 
infinite range interaction from the single light mode of the cavity that couples all atoms with strength $g$. 
The atoms can spontaneously decay with rate $\gamma$ and photons can leave the system through 
the cavity mirrors with rate $\kappa$. 
The system is driven from the side with a laser of Rabi frequency $\Omega$ to deposit excitations 
into the system. Driven dipoles in other lattice geometries are also interesting to consider \cite{bettles15}.}
\label{Fig:ExperimentalSetUp}
\end{figure}

As a suitable model (Fig.~\ref{Fig:ExperimentalSetUp}), we propose to supplement the existing experimental 
set-ups \cite{landig16, klinder15} by weakly admixing a Rydberg-level with relatively low 
principal quantum number ($n\sim 30$).
The other, but equivalent, point of view is to couple a Rydberg-dressed spin lattice 
\cite{zeiher16} to a single mode of an optical resonator. See also 
Refs.~\onlinecite{boddeda16} for a related setup but without a lattice.
As we derive below in Sec.~\ref{sec:model}, the pure spin-part of the Hamiltonian 
$H = H_{\rm spin} + H_{\rm spin-light}$ is
\begin{align}
H_{\rm spin} = -\frac{\Delta}{2}\sum_{\ell=1}^{N}\sigma^z_{\ell}+\frac{V}{d}\sum_{\braket{\ell m}}
\left(\frac{1+\sigma^z}{2}\right)_{\ell}
\left(\frac{1+\sigma^z}{2}\right)_{m}\;,
\label{eq:H_spin}
\end{align}
where the sum $\sum_{\langle \ell m\rangle}$ goes over all nearest-neighbor 
pairs of the square lattice and $d=z/2$ is the dimension of the lattice with $z$ the coordination number. 
For a negative ($-\Delta < 0$) longitudinal field it is favorable for the spins to
point up along the $z$-axis $|\uparrow \rangle$. Competing against this 
is the repulsive or antiferromagnetic ``Rydberg-mediated'' term, which minimizes 
energy by pushing the spin in spatially alternating configurations, e. g. 
$|\uparrow \downarrow \uparrow \downarrow ... \rangle$. In contrast to a conventional Ising $\sim \sigma^z_{\ell} \sigma^z_{m}$ interaction term, the Rydberg interaction is conditioned on population in the upper state.

Non-trivial quantum fluctuations are added to $H_{\rm spin}$ by coherent 
conversion of spin excitations into photons with rate $g$
\begin{align}
H_{\rm spin-light} = 
\frac{g}{\sqrt{N}}(a+a^{\dagger})\sum_{\ell=1}^{N}(\sigma^+_{\ell}+\sigma^-_{\ell})
+
\omega_0 a^{\dagger}a\;,
\label{eq:H_spin-light}
\end{align}
where $\omega_0$ is the effective cavity frequency in a rotating frame and $N$ 
is the number of atoms; the rescaling of the effective spin-light coupling 
with $1/\sqrt{N}$ ensures a non-trivial thermodynamic limit (i.e.\,taking the 
system size and number of atoms to infinity keeping the atomic density and 
electric field strength per volume constant). All coupling constants 
appearing here are expressed in terms of fundamental quantum-optical parameters in 
Sec.~\ref{sec:model} and $g$ is proportional to the external laser drive $\Omega$. 
Note that the $(a+a^\dagger)$ may be viewed to act on the spins like a ``transverse field''
in $x$-direction, whose value depends on the quantum state of the photons. 
Quantum optically, both, the co- and counter rotating terms appearing in Eq.\,\eqref{eq:H_spin-light} are 
naturally induced by the cavity-assisted Raman transitions, see also Sec.~\ref{sec:model}.
%While the energy-conserving terms describe coherent exchange of spin and light excitations, the counter-rotating 
% terms describe the scattering of photons from the pump-laser into the cavity and vice versa. The latter process is needed to %guarantee that the lossy cavity can maintain a fixed number of excitations in the steady-state. Consequently, 
%a rotating-wave approximation would not be an appropriate description of the system we consider. 
%
Within our mean-field treatment, the light-field in the cavity can either be the vacuum mode or it can be in a coherent state. 
By cavity vacuum, we mean the Fock state with zero photon excitations: $a |0\rangle = 0$. Then, 
$\langle a + a^\dagger \rangle = 0$ and the spins 
see zero transverse field. If there is macroscopic occupation of the cavity mode, then $\langle a + a^\dagger \rangle \neq 0$ and the system is in a superradiant state. 

Our model is completed by the inclusion of Lindblad operators for photon losses through the 
mirrors with rate $\kappa$ and spontaneous emission of the atoms with rate $\gamma$ into 
the reservoir modes of the electromagnetic vacuum surrounding the cavity:
\begin{align}
\mathcal{L}_\gamma[\rho]&=\frac{\gamma}{2}\sum\limits_{\ell=1}^N\bigg[2 \sigma^{-}_{\ell} \rho \sigma^{+}_{\ell}-\{\sigma^{+}_{\ell}\sigma^{-}_{\ell},\rho\}\bigg]\label{EQ:LindbladOperatorgamma},\\
\mathcal{L}_\kappa[\rho]&=\kappa\bigg[2 a \rho a^{\dagger}-\{a^{\dagger}a ,\rho\}\bigg]\;,
\label{EQ:lindbladphoton}
\end{align}
where $\rho$ is the system density matrix. 
Spatially modulated phases in the presence of coherent driving of lattice atoms have 
been discussed in an open, non-equilibrium setting in particular by Lee and collaborators 
\cite{tony11,tony13}; see also Ref.~\onlinecite{hoening14}. Here, we extend such models by coupling the spin degrees of freedom to a quantum light field, which can also be in a zero-photon vacuum state with undetermined phase. 
In fact, the loss rate for the photons $\kappa$ wants to drive the photons into this vacuum state (i.e.\,the empty cavity 
$| 0 \rangle$).

Quite generally, driven dissipative lattice models are currently under intense investigation. The systems range from (effective) spin-1/2 \cite{tony11,tony13,Angelakis16,WangLuo16} and Bose-Hubbard \cite{Boite13,Boite14} models to systems with interacting photons in cavity arrays \cite{Zou14,Wilson16,Schiro16,YuNaZhang14}. 
%The superradiance transition of a related model of an Ising chain in a lossy cavity was recently analysed \cite{WangLuo16}. Earlier investigations also treated superradiance transitions in a closed system of a spin chain with nearest-neighbour interactions coupled to a multimode optical resonator \cite{YuNaZhang14}.

%We mention that if the atom-light coupling was replaced with a coherent driving term with Rabi-frequency $\Omega$, such that $\frac{g}{\sqrt{N}}(a+a^{\dagger})\sum_{\ell=1}^N\left(\sigma^{+}_{\ell}+\sigma^{-}_{\ell}\right)\to\frac{\Omega}{2}\sum_{\ell=1}^N \left(\sigma^{+}_{\ell}+\sigma^{-}_{\ell}\right)$ our model reproduces the simpler dynamics 
%(without photon losses) of Ref.~\onlinecite{tony11}. 

We now present our main results from an analysis 
of Eqs.~(\ref{eq:H_spin}-\ref{EQ:lindbladphoton}) using even-odd sublattice 
mean-field master equations (derived in Sec.~\ref{sec:langevin}) and the 
input-output formalism. A detailed discussion and derivation of these results 
can be found in Sec.~\ref{sec:results}.

\subsection{Result 1: Combination of superradiance and magnetic translation symmetry-breaking}
\label{intro:result1}

\begin{figure}
\includegraphics[width=110mm]{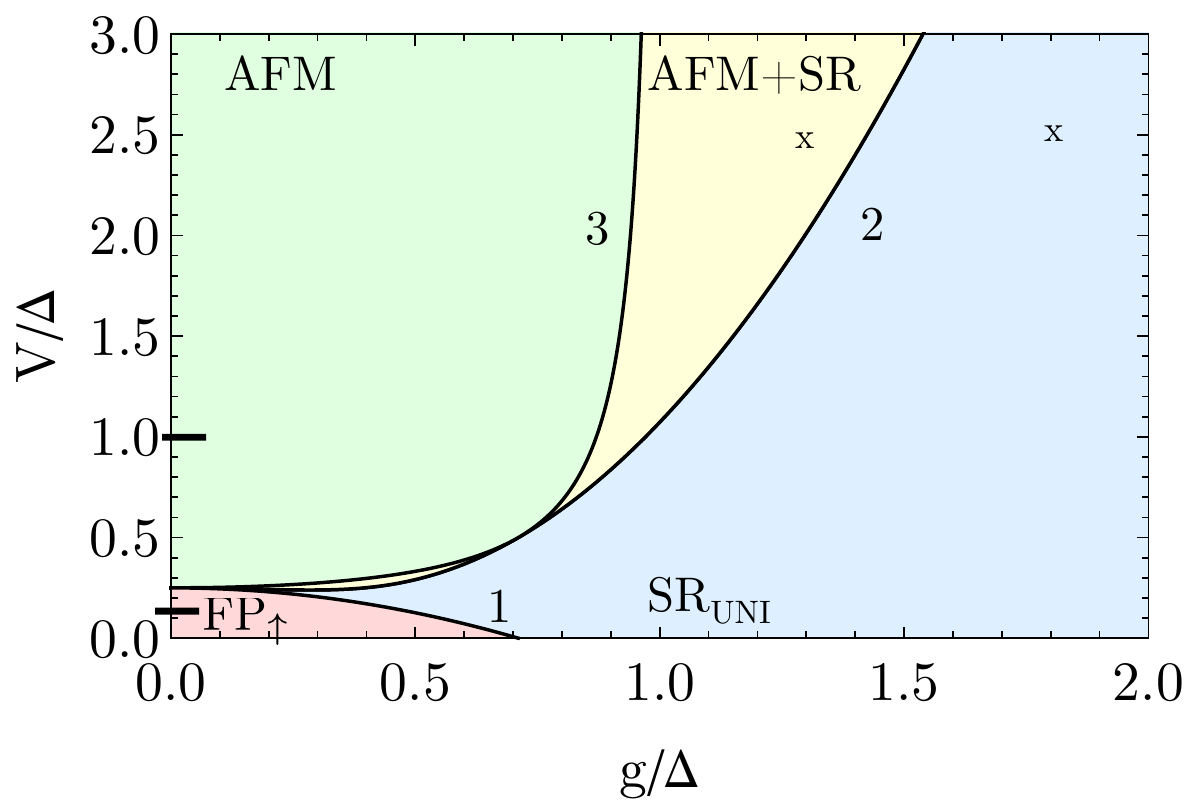}
\caption{Non-equilibrium mean-field phase diagram of a Rydberg-dressed spin 
lattice (with nearest neighbor interaction $V$) coupled to a single-mode optical light 
field with rate $g$. In units of the spin longitudinal field $\Delta$. 
The key feature is the yellow strip, AFM+SR phase, in which 
spatially modulated magnetic moments occur together with a superradiant photon condensate, see 
Tab.\,\ref{Tab:OrderParameters}.
This phase may be regarded as the magnetic analogue of the 
(superradiant) supersolid of moving lattice bosons in an optical cavity \cite{landig16,dogra16,chen16,li13}.
The yellow strip merges into the $V$-axis at a multi-critical point from which four different phases can 
be reached by infinitesimal variation of parameters. At the multi-critical point 
$(g/\Delta = 0, V/\Delta = 1/4)$ the spins are ``maximally soft'', i. e., 
they feel zero effective, longitudinal field.
Here this infinitesimal sensitivity to ordering is not rounded off by cavity decay induced noise. 
Recall that typically the cavity-induced decay rate for an atom is 
$\sim g^2\kappa/(\omega_0^2 + \kappa^2)$.
Here however the noise kicks of the two photon 
absorption and emission pathways destructively interfere in the level scheme 
Fig.~\ref{Fig:RydbergDressingScheme}, which, in turn, 
manifests itself in the $\sim g(\sigma_\ell^+ + \sigma_\ell^-)(a+a^\dagger)$ coupling.
SR$_{\rm UNI}$ is a uniform superradiant phase in which the spins also develop 
an expectation value in $x$-direction. AFM stands for antiferromagnetic with 
differing magnetic moments on the even and the odd sublattice.
FP$_{\uparrow}$ is a fully polarized phase in which all spins point 
up. The magnetisations and the value of the photon condensate across the transitions are continuous. 
Cavity spectra at positions labeled with (x) are depicted in Fig.\,\ref{Fig:Cavity Spectra}. 
Numerical parameters used: $\omega_0/\Delta=2.0, \kappa/\Delta=0.2$.}
\label{Fig:DRPhaseDiagramkappaonly}
\end{figure}

Our first key result is Fig.~\ref{Fig:DRPhaseDiagramkappaonly}: the non-equilibrium steady-stase phase diagram of Eqs.~(\ref{eq:H_spin}-\ref{EQ:lindbladphoton}) setting the atomic spontaneous emission $\gamma$ to zero for now. 
Using cavity-assisted Raman transitions \cite{dimer07} to tune the atom-light coupling, this describes the limit of 
relatively far detuned excited states, where population in 
the decaying levels is suppressed. 
These phase diagrams are computed from solving for steady states of mean-field master equations 
for the real-valued atomic variables $(\braket{\sigma^x},\braket{\sigma^y},\braket{\sigma^z})$ and the 
complex-valued photon expectation values $(\braket{a},\braket{a^{\dagger}})$, see Eqs.~(\ref{EQ:Sigmax1}-\ref{EQ:PhotonTerm}).

This way of solving the problem implicitly 
takes first the thermodynamic limit $N\rightarrow \infty$ and subsequently the long-time limit $t\rightarrow\infty$. 
We keep $\kappa$ finite to account for photon losses. In App.~\ref{Sec:Equilibrium}, we show how the somewhat unphysical limit $\kappa \rightarrow 0$ reproduces in fact the phase boundaries of a corresponding 
ground state $T=0$ model. 
In a quantum optics experiment, this most closely seems to correspond to the 
preparation protocol in which the interacting spins are prepared in a low-entropy state in a given phase, first without 
any coupling to the cavity (i.e.\,the transversal laser drive $\Omega$ turned off).
Starting in the AFM phase in Fig.~\ref{Fig:DRPhaseDiagramkappaonly}, for example, 
the coupling $g$ is then turned on to induce superradiance 
and the AFM+SR phase. However, 
existing experiments and in particular the onset of superradiance are surprisingly robust 
against variations in the preparation scheme \cite{baumann10,klinder15,landig16}.

%In principle, the ramp of $g$ should be sufficiently 
%slow not to de-equilibrate the AFM state but fast enough compared to $1/\kappa$ to avoid loosing
%photons from the cavity before they can interact and mediate the spin interaction. 

The phases shown in Fig.~\ref{Fig:DRPhaseDiagramkappaonly} 
can be classified according to their ``order parameters'' in Table \ref{Tab:OrderParameters}.
\begin{table}
\begin{center}
\begin{tabular}{ c | c | c }
Phase & Broken Symmetry
& Order Parameter \\ \hline
SR$_{\rm UNI}$ & 
Superradiance $\mathbbm{Z}_2$
& $\braket{a}\neq 0$\\
AFM & Lattice translations $T_{\rm lat}$ & $\braket{\sigma^z_e}-\braket{\sigma^z_0}\neq 0$\\ 
AFM+SR &   $\mathbbm{Z}_2$ {\emph{and}} $T_{\rm lat}$ & $\braket{\sigma^z_e}-\braket{\sigma^z_0}\neq 0$, $\braket{a}\neq 0$
\\
FP & None & None
\end{tabular}
\caption{
Order parameters for the phases in Fig.~\ref{Fig:DRPhaseDiagramkappaonly}. Whenever 
the photon parity is broken, the $x$-projections of the spins also attain a finite expectation value 
$\langle \sigma^x \rangle \neq 0$.}
\label{Tab:OrderParameters}
\end{center}
\end{table}
Let us describe the phases in more detail. Upon increasing the coupling to the photons 
along the $g$-axis in Fig.~\ref{Fig:DRPhaseDiagramkappaonly}, 
for $V/|\Delta| < 1/4$, a fully polarized phase (FP$_{\uparrow}$, $|\uparrow \uparrow ... \rangle$) 
becomes superradiant crossing the  
Dicke transition, which has been studied in detail for both, the closed thermal and ground states 
as well as the open variant (most recently including also single-site atomic spontaneous 
emission \cite{baden14,gelhausen16}). The symmetry, which is spontaneously broken is
\begin{align}
\mathbbm{Z}_2: [a+a^{\dagger},\sigma^x_{\ell},\sigma^y_{\ell}]\to [-(a+a^{\dagger}),-\sigma^x_{\ell},-\sigma^y_{\ell}],
\label{eq:Z2}
\end{align}
The experimental signature is a jump of the photon number 
inside the resonator \cite{landig16,baden14}. Recall that here we have the atomic spins 
pinned in an optical lattice, which takes away the photon recoil momenta transferred 
from the photons to the atoms. This is in contrast to the realizations of the Dicke model 
with momentum states of the atomic gas \cite{domokos02,black03,baumann10}; therein 
the onset of superradiance is accompanied by even-odd checkerboard formation. 
Here, with the setup Fig.~\ref{Fig:ExperimentalSetUp}, the superradiance leads 
to uniform spin polarization in $x$-direction and the Rydberg-mediated repulsion 
{\emph{competes}} with this and tends to break the even-odd lattice translation 
symmetry.

Up the $V$-axis, at $g=0$ the magnetisations of the system can change discontinuously (at $V=\Delta/4$ in two dimensions) from a fully polarised state (FP$_{\uparrow}$) to an antiferromagnetic 
excitation pattern (AFM). This AFM phase breaks a discrete even-odd translation symmetry
\begin{align}
T_{\rm lat}:[\sigma^{\alpha}_{e,o}]\to [\sigma^{\alpha}_{o,e}]\;,
\end{align}
which can lead to different sublattice magnetisations as depicted in Fig.\,\ref{Fig:DRPhaseDiagramkappaonly}. Here, $T_{\rm lat}$ exchanges the even (e) and odd (o) sublattice index of the atomic variables $\sigma^{\alpha}$ with $\alpha=\{x,y,z\}$ in the Hamiltonians in Eqs.\,(\ref{eq:H_spin}-\ref{eq:H_spin-light}). 
As described further in the caption of Fig.~\ref{Fig:DRPhaseDiagramkappaonly}, 
the AFM+SR phase has a curious feature, namely that it is split in two regions, that 
are delimited by a touching point of two second-order phase transition lines of the ${\rm SR_{UNI}}$ 
and the plain AFM phase at $V/|\Delta|=1/2$. For this special value, the effective magnetic field 
on one sublattice vanishes. At this point, however, the transition becomes discontinuous. 
Direct, continuous transitions between phases with different broken symmetries are rare and 
sometimes accompanied by deconfined quantum criticality, as in the 
ground state of frustrated quantum spin models, for example \cite{senthil04}.

\subsection{Result 2: Even-odd sublattice peak in cavity spectrum}
\label{intro:result2}

A convenient feature of these quantum-optical quantum simulators is the ability to 
perform non-destructive measurements of the dynamics via cavity 
spectra. For the moving atomic quantum gas and the associated polaritonic 
density excitations, such measurements have lead to fruitful 
experiment-theory comparison \cite{mottl12,brennecke13,piazza14b,kulkarni13,konya14}.

\begin{figure}
%\subfloat[]{\includegraphics[width=8cm]{CavSpectrTrivAF.pdf}\label{subfig:CavSpectrTrivAF}}
\subfloat[]{\includegraphics[width=10cm]{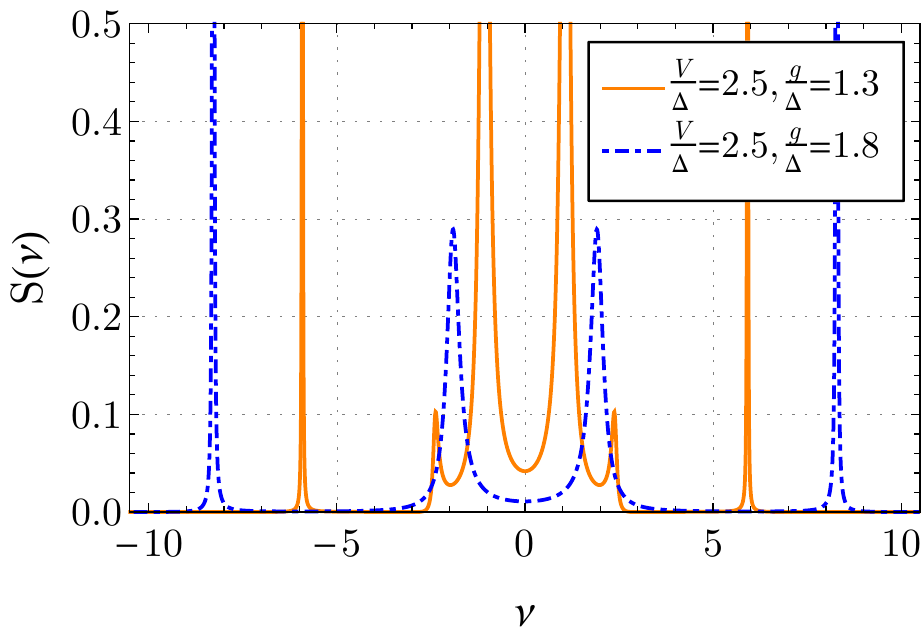}}
\subfloat[]{\includegraphics[width=6cm]{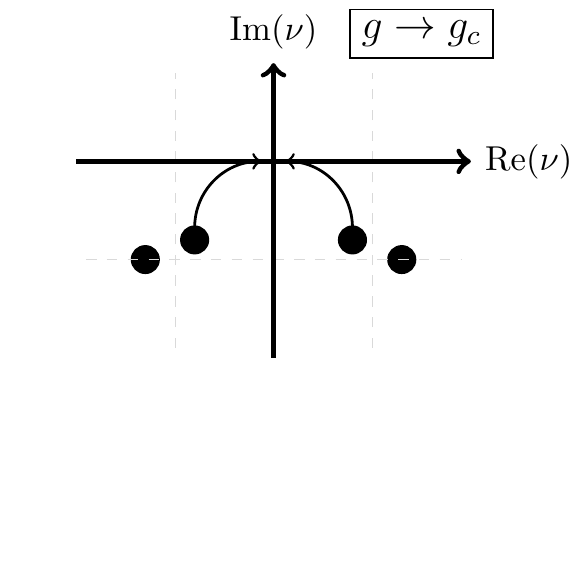}\label{subfig:DickeRydbergPoles}}
%\subfloat[]{}
%\begin{tikzpicture}
%\filldraw (-.8,-.8) circle (4pt);   %critical pole
%\filldraw (.8,-.8) circle (4pt);    % critical pole
%\filldraw (-1.3,-1.) circle (4pt); % spectator pole
%\filldraw (1.3,-1.) circle (4pt); % spectator pole
%    \node[font=\fontsize{12}{24}\selectfont,draw] at (1.5,1.3) {$g \to g_c$}; % label
%\draw[help lines, color=gray!30, dashed] (-1.9,-1.9) grid (1.9,0.9);% coordinate system 1/3
%\draw[->,ultra thick] (-2,0)--(2,0) node[right]{Re$(\nu)$}; % coordinate system 2/3
%\draw[->,ultra thick] (0,-2)--(0,1) node[above]{Im$(\nu)$}; % coordinate system 3/3
%\path (-0.8,-0.8) node(x) {} % path of critical pole 1/3
%(0,0) node(y) {};% path of critical pole 2/3
%\draw[->,thick,black] (x) .. controls +(up:0.5cm) and +(left:0.5cm) .. node[above,sloped] {} (y); %path of critical pole 3/3
%\path (0.8,-0.8) node(x) {} % path of critical pole 1/3
%(0,0) node(y) {};% path of critical pole 2/3
%\draw[->,thick,black] (x) .. controls +(up:0.5cm) and +(right:0.5cm) .. node[above,sloped] {} (y); %path of critical pole 3/3
%\end{tikzpicture}
\caption{Even-odd sublattice peak in the cavity spectrum (peak close to zero frequency of the orange, solid line), 
which appears when the translational lattice symmetry 
$T_{\rm lat}$ is spontaneously broken by the Rydberg-dressed nearest-neighbor interaction $V$.
The two cavity spectra are computed for the positions labeled by (x) in Fig.\,\ref{Fig:DRPhaseDiagramkappaonly}. The blue, dashed line has the two polariton peaks in the uniform SR$_{\rm UNI}$ phase 
with the photonic branch around the cavity frequency $\omega_0/\Delta=2.0$. 
The orange, solid line is the spectrum in the AFM+SR regime with broken $T_{\rm lat}$; it shows 
the prominent even-odd peak, which becomes soft toward the phase boundary (the right 
edge of the yellow strip in Fig.~\ref{Fig:DRPhaseDiagramkappaonly}. 
(b) Low-energy pole structure of the even-odd sublattice peak, where both the real and the imaginary part of the poles vanish simultaneously as $g\to g_{c}$ according to 
Eq.~(\ref{eq:powers}).}
\label{Fig:Cavity SpectraIC}
\end{figure}

Here we show that the translation-symmetry breaking induced 
by the nearest-neighbor Rydberg-dressed interaction $V$ leads to a novel 
collective mode and peak in the spectrum. 
Figure \ref{Fig:Cavity SpectraIC} shows the cavity spectrum upon increasing 
$g$ from the (AFM+SR) phase with broken $T_{\rm lat}$ symmetry into the SR$_{\rm UNI}$ phase where translation symmetry is restored.
We observe that the conventional normal-mode polariton picture 
first seen by Rempe and Kimble \cite{thompson92} --which is, with modifications, also
applicable to the Dicke model \cite{dalla13}-- becomes insufficient 
to describe the photon dynamics. By contrast, the even-odd mode softens 
in a specific way: Denoting the frequency of the polariton pole as $\nu$, we observe that both, its real part (gap) and the imaginary part (damping rate)
vanish with coupling constant (from the AFM+SR phase toward line 2 
in Fig.~\ref{Fig:DRPhaseDiagramkappaonly}) with 
\begin{align}
{\rm Re}[\nu]\propto \pm\sqrt{g_{c}-g}, \quad{\rm Im}[\nu]\propto -(g_{c}-g)\;,
\label{eq:powers}
\end{align}
where $g_c$ refers to the right boundary delimiting the AFM+SR phase. This is because the translational symmetry $T_{\rm lat}$ affects the atomic sector only which does not couple directly to the photonic rate of dissipation $\kappa$. This is in contrast to Dicke-type models
or superfluids out-of-equilibrium (see \cite{szymanska06} and \cite{szymanska07})
whose dynamics becomes purely overdamped/imaginary at the transition 
\cite{dalla13}, that is, the real part of the mode vanishes first. 

\subsection{Result 3: Photon number oscillations}
\label{intro:result3}

We now account for a non-zero rate of atomic spontaneous emission $\gamma \neq 0$. 
Specific details of a given quantum-optical implementation (see Sec.~\ref{sec:model})
will determine which set of Lindblad operators and additional atomic levels 
need to be accounted for. In order to gain a first qualitative picture, we model an {\emph{effective}} 
decay rate with $\mathcal{L}_\gamma[\rho]$ in Eq.~(\ref{EQ:LindbladOperatorgamma}) between 
the effective spin-up and spin-down states ($|1\rangle$ and $|0\rangle$ 
in Fig.~\ref{Fig:RydbergDressingScheme}. We expect 
$\gamma$ to become larger once the detuning to the shorter lived excited 
states is decreased; it is generally true that the effective ground state levels inherit 
a finite lifetime from admixing a short-lived state. For a specific experimental set-up, 
one may also include other types of atomic losses or dephasing.

\begin{figure}
\subfloat[]{\includegraphics[width=75 mm]{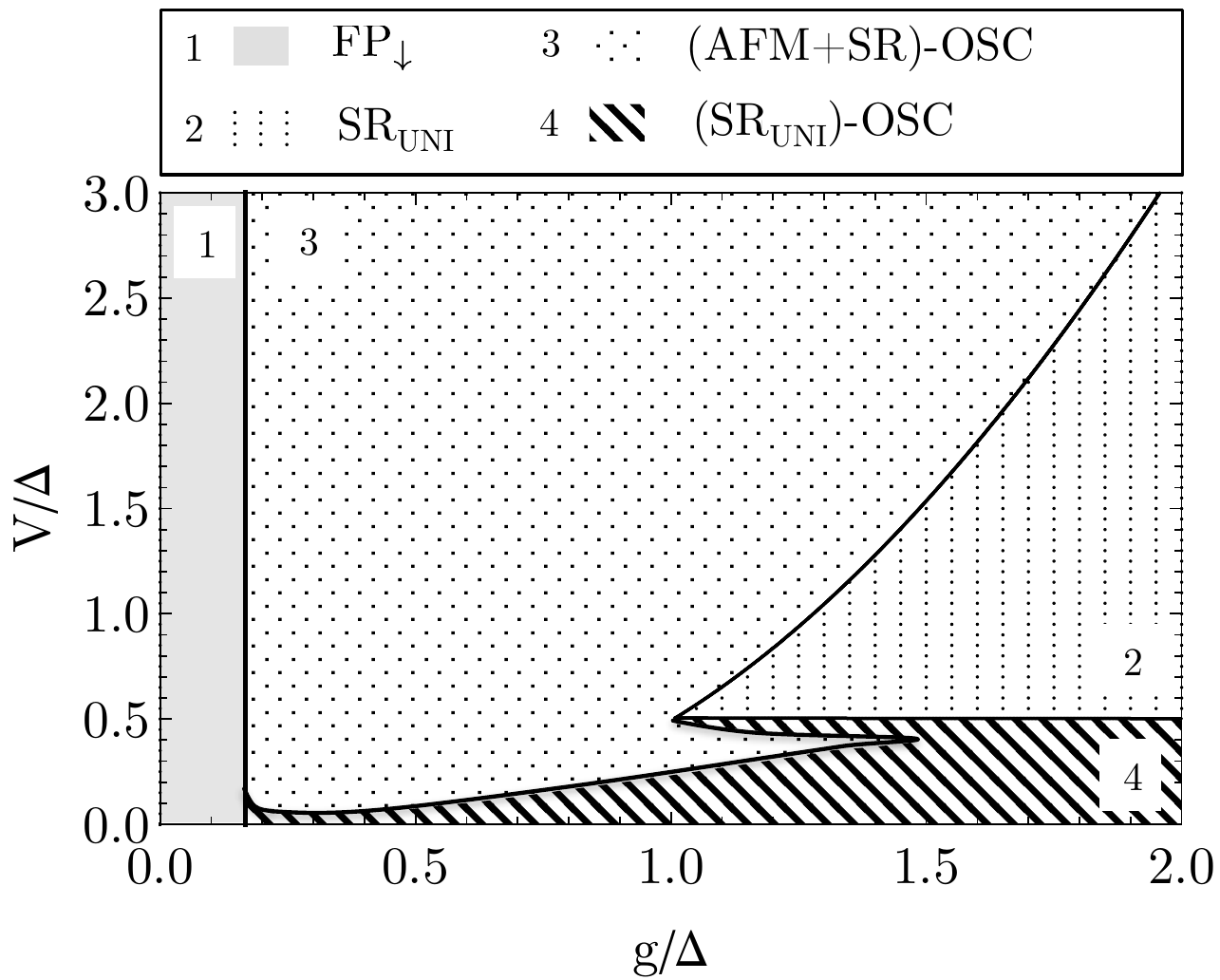}\label{subfig:PhaseDiagramVGsmallgamma}}
\subfloat[]{\includegraphics[width=90mm]{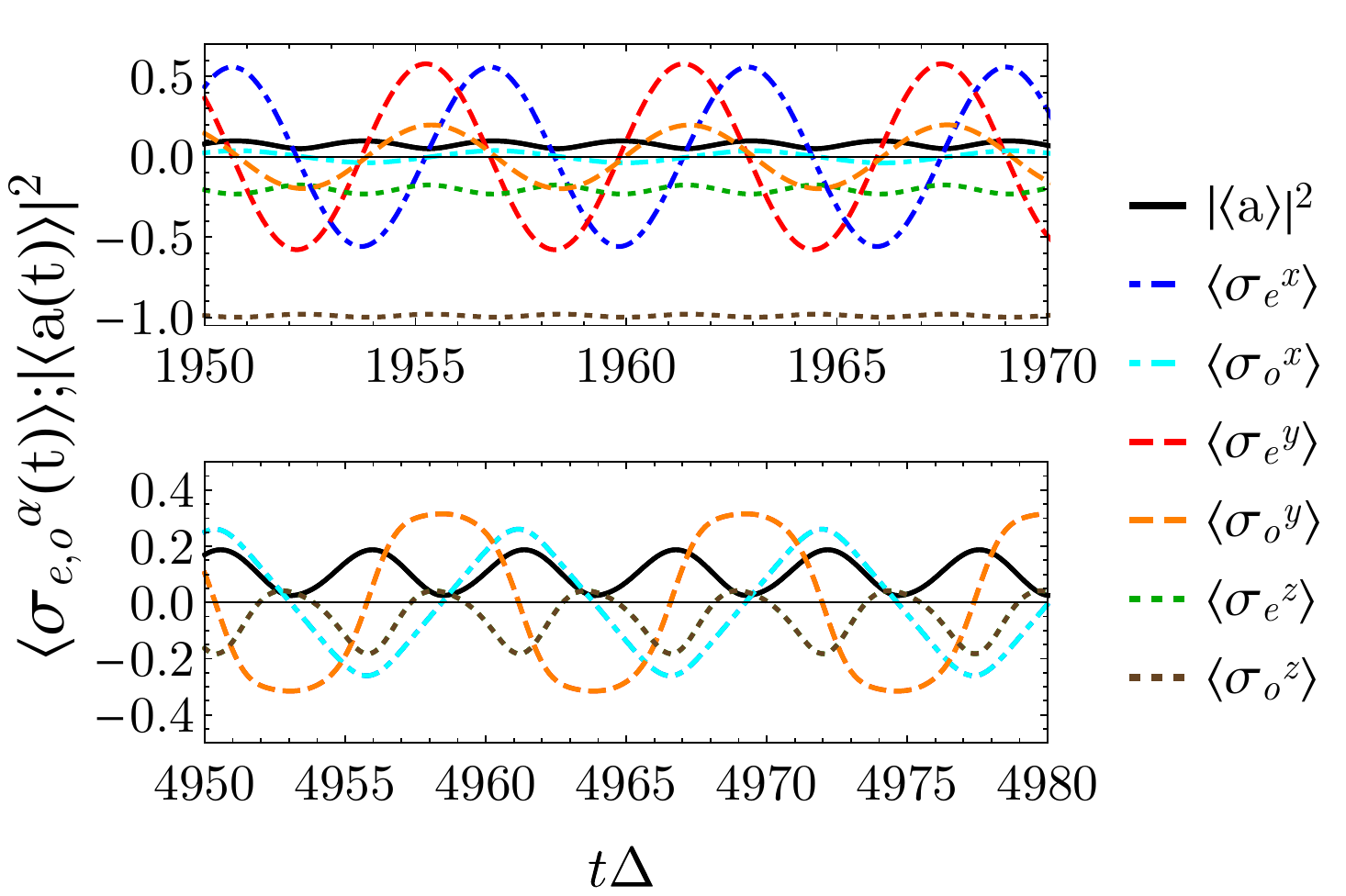}\label{subfig:OscillationsVGPhases}}
\caption{(a) Supplementing the phase diagram Fig.~\ref{Fig:DRPhaseDiagramkappaonly} 
in the (V/$\Delta$,g/$\Delta$)-plane by a small amount of atomic dissipation $\gamma/\Delta=0.01$ with $(\omega_0/\Delta=2.0,\kappa/\Delta=0.2)$ yields a different picture. In comparison to the $\gamma/\Delta=0.0$ case (compare Fig.\,\ref{Fig:DRPhaseDiagramkappaonly}), there are no stable steady-states with a broken lattice symmetry $T_{\rm lat}$ any more. Instead, the system can show persistent oscillations in time. (b) Depiction of the 
non-uniform (top, (AFM+SR)-OSC) oscillations and the 
uniform (bottom, SR$_{\rm UNI}$-OSC) oscillations that characterize the long time limit behavior of Eqs.~(\ref{EQ:Sigmax1}-\ref{EQ:PhotonTerm}) with finite $\gamma$. Parameters for the upper plot are $(V/\Delta=g/\Delta=0.5)$, the lower plot is obtained at $(V/\Delta=0.3, g/\Delta=1.2)$.}
\label{FIG:VGPhasesandOscillations}
\end{figure}

This at first sight innocuous change has interesting consequences. 
Even qualitative features of Fig.~\ref{Fig:DRPhaseDiagramkappaonly} 
are drastically changed (although experimentally, for far enough detuned 
intermediate, excited states and rapid, enough ramps of the spin-light 
coupling $g$ these $\gamma$-induced 
changes may not be immediately visible).
Allowing for a small $\gamma$, see Fig.~\ref{FIG:VGPhasesandOscillations}, 
in particular wipes out the stable AFM phase and introduces 
a fully downward polarized state FP$_{\downarrow}$ as well as a novel
oscillatory phase (AFM+SR)-OSC.  Here also the 
photon field amplitude oscillates which can be detected by time-resolved measurements of the 
intensity of the light leaving the cavity. 

At the root of these effect is $\mathcal{L}_\gamma[\rho]$ in Eq.~(\ref{EQ:LindbladOperatorgamma}): it explicitly 
breaks the discrete symmetry $G$ given by the product of time-reversal: $\mathcal{T}_{\ell}=-i\sigma^y_{\ell}K_\ell, \ t\rightarrow-t$ (for a spin $s=1/2$) 
and spin rotation around the y-axis by $\pi$: $D^{1/2,\ell}_{y,\pi}=-i\sigma^y_{\ell}$. Here $K_\ell$ is the 
complex conjugation operator such that 
$\mathcal{G}_{\ell}=D^{1/2,\ell}_{y,\pi} \mathcal{T}_{\ell}=-K_{\ell}$. 
If we write $G=\Pi_{\ell=1}^N\mathcal{G}_{\ell}$ we have
$GHG^{-1}=H$. In particular, this implies for steady states 
 $\langle\sigma^y_{e,o}\rangle=\langle G\sigma^y_{e,o}G^{-1}\rangle=-\langle\sigma^y_{e,o}\rangle\overset{!}{=}0$ if $\gamma = 0$. 
 For $\gamma \neq 0$, the spins can start developing expectation values also in the $y$-direction.
 This offers new possibilities for the spin dynamics such as 
 anomalous spin precession \cite{tony13} not available in equilibrium.
 
%This at first sight innocuous change has interesting consequences. This 
%is because $\mathcal{L}_\gamma[\rho]$ in Eq.~(\ref{EQ:LindbladOperatorgamma}) explicitly 
%breaks the discrete symmetry $G$ given by the product of time-reversal 
%and spin rotation around the y-axis by $\pi$: 
%$\mathcal{T}_{\ell}=-i\sigma^y_{\ell}K_\ell, \ t\rightarrow-t$ (for a spin $s=1/2$) 
%and rotation in spin-space around the $y$-axis with angle $\theta=\pi$ 
%denoted as $D^{1/2,\ell}_{y,\pi}=-i\sigma^y_{\ell}$, where $K_\ell$ is the 
%complex conjugation operator such that 
%$\mathcal{G}_{\ell}=D^{1/2,\ell}_{y,\pi} \mathcal{T}_{\ell}=-K_{\ell}$ and 
%$\mathcal{G}_{\ell}\mathcal{G}^{-1}_{\ell}=1$ with $[\mathcal{G}_{\ell},\mathcal{G}^{-1}_m]=0$ 
%and $[\mathcal{G}_{\ell},\mathcal{G}_m]=0$. 
%If we write $G=\Pi_{\ell=1}^N\mathcal{G}_{\ell}$ we have
%$GHG^{-1}=H$. In particular, this implies for steady states 
% $\langle\sigma^y_{e,o}\rangle=\langle G\sigma^y_{e,o}G^{-1}\rangle=-\langle\sigma^y_{e,o}\rangle\overset{!}{=}0$ if $\gamma = 0$. 
% For $\gamma \neq 0$, the spins can start developing expectation values also in the $y$-direction.
% This offers new possibilities for the spin dynamics such as 
% anomalous spin precession \cite{tony13} not available in equilibrium. 
 
 %
\begin{figure}
\includegraphics[width=180mm]{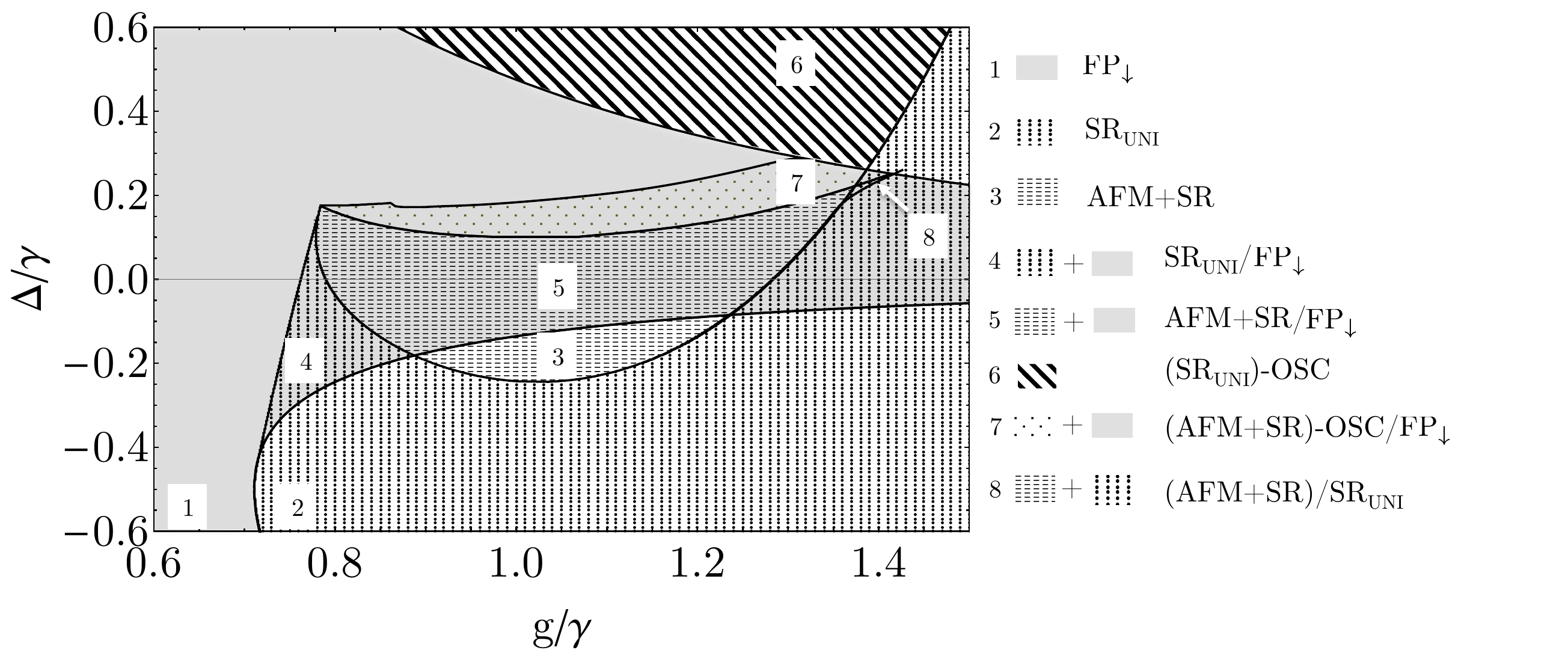}\label{subfig:DRPhaseDiagramkappagamma}
%\subfloat[]{\includegraphics[width=8cm]{MagneticOrderingswithgamma2.pdf}\label{subfig:OrderParameterskappagamma}}\\
\caption{Non-equilibrium steady-state phase diagram of Eqs.~(\ref{eq:H_spin}-\ref{EQ:lindbladphoton}) 
with finite, atomic spontaneous emission $(\kappa/\gamma=0.2, V/\gamma=1.8,\omega_0/\gamma=2.0)$.
Apart from time-independent states, the dynamics also realises limit cycles where atomic and photon components show persistent oscillations in time. Oscillations can be uniform or different on the even/odd sublattice, see Fig.\,\ref{fig:oscillations}. Depending on the initial configuration, the system can reach different long-time fixpoints. Bistabilities occur whenever two phases overlap (see legend). Crystalline antiferromagentic order only occurs together with superradiance (AFM+SR).}
\label{Fig:DRPhaseDiagramkappagamma}
\end{figure}

We show the phases for a further range of parameters $(\Delta/\gamma ,g/\gamma)$ space for a fixed strength $V$ of the Rydberg interaction in two spatial dimensions in Fig.~\ref{Fig:DRPhaseDiagramkappagamma}. In mean-field theory we distinguish five phases in the long time limit. Three are steady-states denoted as (FP$_{\downarrow}$, ${\rm SR_{UNI}}$, AFM+SR, see also Tab.\, \ref{Tab:OrderParameters}) and two are stable limit cycles. 
The Lindblad operators try to drive the system into an empty state without any excitations; consequently 
AFM order can only occur in the presence of a coherent drive, i.e.\, together with a photon condensate $\braket{a}$. In the latter phases, the system exhibits oscillations in both atomic and photonic components, since the atomic dynamics couples back to the photon sector through Eq.\,\eqref{EQ:Dickea}. The oscillations can be uniform in all components ${\rm(SR_{UNI})-OSC}$ or different on the sublattices ${\rm (AFM+SR)-OSC}$. 

 \begin{figure}
\subfloat[]{\includegraphics[width=125mm]{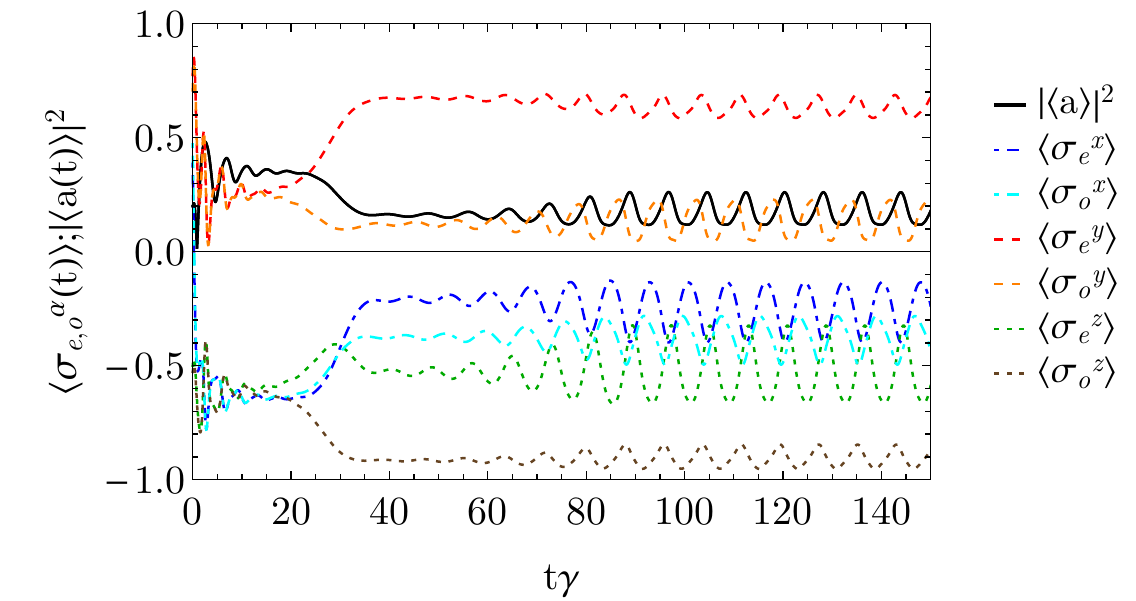}}
\subfloat[]{\includegraphics[width=5cm]{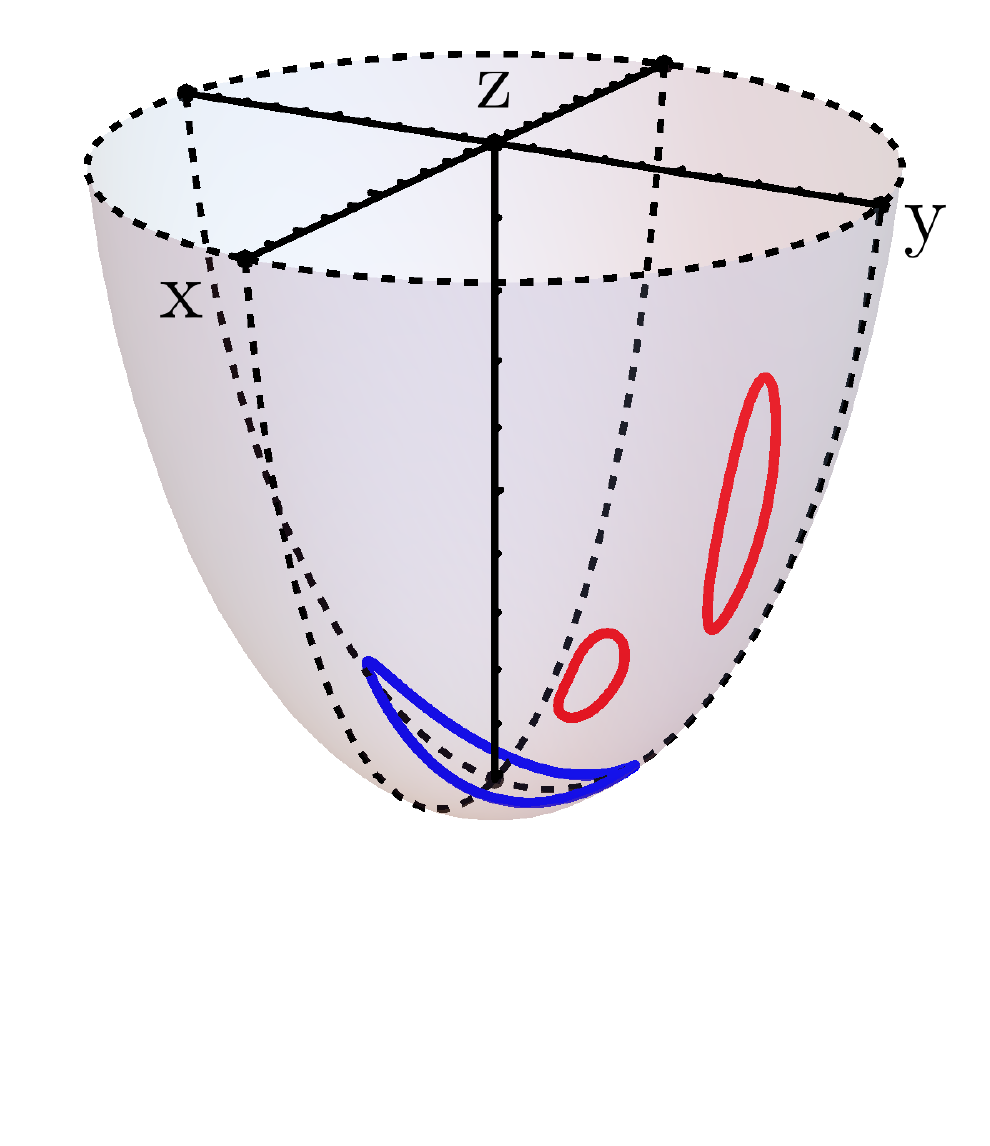}}
\caption{(a): Persistent spin and photon-field oscillations appear in the (AFM+SR)-OSC phase for $\Delta>0$ 
close to the (AFM+SR) stable region for the parameters used in Fig.\,\ref{subfig:DRPhaseDiagramkappagamma} for $g/\gamma=1.1,\Delta/\gamma=0.15$. 
Note that between $t\gamma \approx 10 - 25 $ the amplitudes display 
(quasi-) plateaus followed by a rapid in/decrease toward their final mean values; and 
only then the oscillations begin and persist. For moving atomic gases 
in a cavity, we note that Ref.~\onlinecite{schutz14} found 
``pre-thermalized'' plateaus  in the time evoluation for the order parameter 
by solving Fokker-Planck type kinetic equations.
(b): Illustration of the atomic components for the limit cycles on the lower part of the Bloch sphere. The two upper lineshapes (red) depict the lines traced by the oscillations on the even and the odd sublattice in the (AFM+SR)-OSC phase, as depicted in (a). The lower lineshape illustrates a limit cycle of uniform oscillations of the atomic components in the ${\rm SR_{UNI}}$ phase.}
\label{fig:oscillations}
\end{figure}

One point of view to interpret the oscillations data in Fig.~\ref{fig:oscillations} is that the (collective) spin oscillator (see also 
 Ref.~\onlinecite{bhaseen12}) and  the cavity oscillator synchronize with each other after a sufficient amount of time 
 (see also Ref.~\onlinecite{tony_cavity13}).
Additionally, the system can show bistabilities, meaning that the eventual fate of the system in the long-time limit depends on the initial conditions. 
%In a variety of phases, the system is bistable with the empty atom-cavity system, here denoted as FP$_{\downarrow}$, where all spins are pointing down. 
However, we also find a small strip in the phase diagram where the system is bistable between ${\rm{AFM+SR}}$ and a ${\rm SR_{\rm UNI}}$ phase.
We will describe the phases more in Sec \ref{subsec:result3}.

\section{Ideas for quantum-optical implementation of the model}
\label{sec:model}
\begin{figure}
\includegraphics[width=160mm]{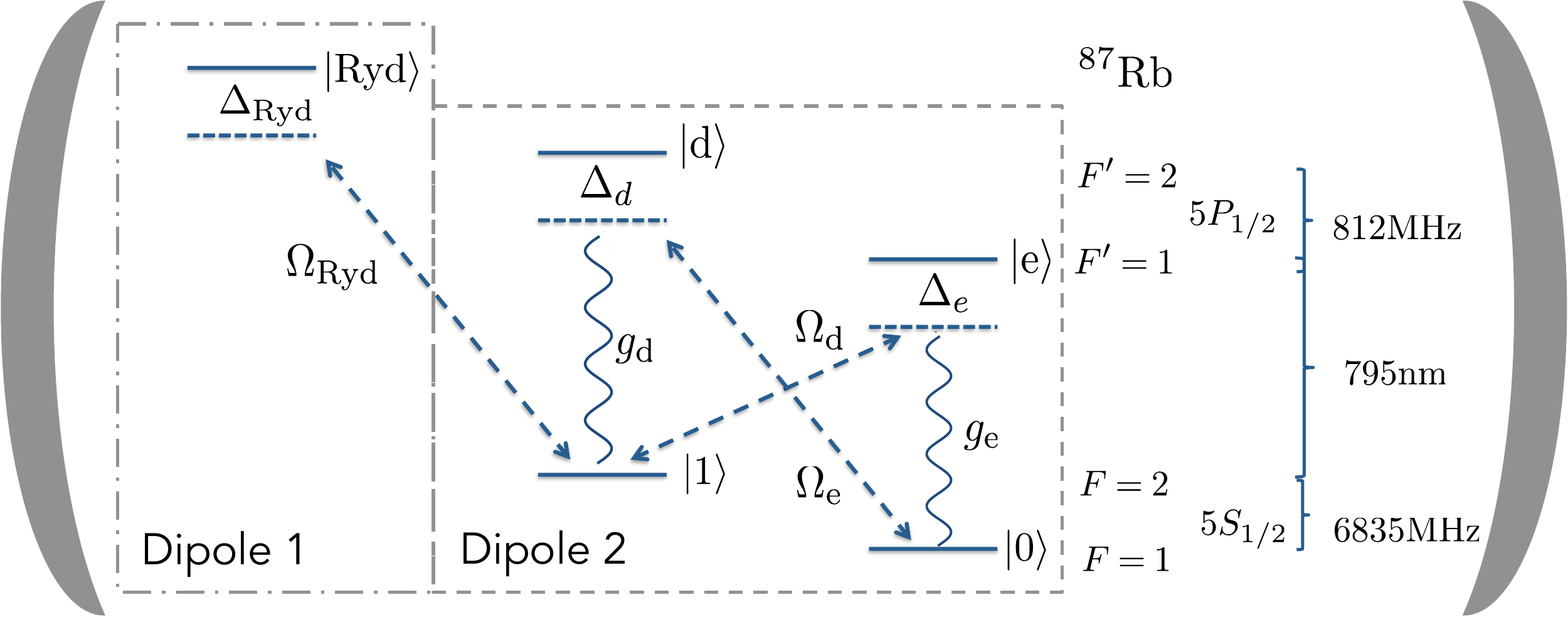}
\caption{ Blueprint for the wanted "two dipoles in-one" unit and the Hamiltonian 
Eqs.~(\ref{eq:H_spin},\ref{eq:H_spin-light}), which 
allows two independent photonic force carriers to couple to the atomic spin.
The effective spin degree of freedom is encoded in the levels $|1 \rangle$ 
and $|0\rangle$. The resonator mirrors (grey shades) confine the optical photon mode, 
which couples to the effective spin via $g_d$ and $g_e$.
{\emph{Dipole 1}}:
%The weak admixing of a Rydberg level to 
%$|1\rangle\rightarrow\ket{1}+\frac{\Omega_{\rm Ryd}}{2\Delta_{\rm Ryd}}\ket{\rm Ryd}$ results 
%in exchange of virtual microwave photons between neighboring sites 
%and induces (virtual) transitions within the Rydberg manifold conditioned on population 
%in the $|1\rangle$ state. 
A dressing laser can weakly admix a Rydberg level to a ground-state $\ket{1}\rightarrow\ket{1}+\frac{\Omega_{\rm Ryd}}{2\Delta_{\rm Ryd}}\ket{\rm Ryd}=\ket{\tilde{1}}$. We want the resulting effective potential $V^{\rm eff}_{\ell m}$ between $\ket{\tilde{1}}_{\ell}$ and $\ket{\tilde{1}}_{m}$ states to be predominantly nearest-neighbor in a square lattice spaced by an optical wavelength. Complex potentials 
including angle-dependencies can be realized \cite{zeiher16,glaetzle14,glaetzle15}. 
{\emph{Dipole 2}}: The double Raman scheme \cite{dimer07} provides 
a tunable coupling $\sim(a+a^\dagger) \sigma^x_\ell$ to the optical cavity.
Choosing the lattice and cavity modefunction as in Fig.~\ref{Fig:ExperimentalSetUp}
results in a homogeneous coupling $g$, that is, all the spins couple in the 
same way to the cavity. This provides the infinite-range coupling between 
all the spins. The depicted level scheme corresponds to the ground state manifold $5 S_{1/2}$ 
and the first excited state manifold $5P_{1/2}$ of ${}^{87}$Rb including their typical frequency splittings.}
\label{Fig:RydbergDressingScheme}
\end{figure}

We seek an implementation which realizes the Hamiltonian in Eqs.~(\ref{eq:H_spin},\ref{eq:H_spin-light}).
At the core of our model is the "two dipoles in-one" unit depicted in 
Fig.~\ref{Fig:RydbergDressingScheme}. As described in the caption, 
Dipole 1 could be created by weakly admixing a relatively low quantum number 
Rydberg level ($n\sim30$) to a set of long-lived hyperfine split states 
$|1 \rangle$ and $| 0 \rangle$. Dipole 2 couples to the cavity via two far-detuned, excited states 
$\ket{d},\ket{e}$.

%
%\,\eqref{eq:hamiltonian} with the conditions that allow for (I) tunable atom-atom and atom-light interactions on the same orders of magnitude, (II) repulsive interactions of Rydberg-dressed atoms on lattice spacings in the optical regime, (III) dominant nearest-neighbour interactions of Rydberg-dressed atoms. 
%Due to their strong scaling properties with the principal quantum number $n$, resonant Rydberg-Rydberg interactions can dominate all other interaction strengths in the system. The bare Rydberg-Rydberg interaction lives on energy scales in the UV-regime (THz) whereas the cavity-mediated Dicke interaction often is in the optical or the MHz regime. An optical pumping scheme must be implemented that bridges the energetic separation between the two competing dipole couplings of atom-light and atom-atom interactions. These two dipoles can be combined within one atomic qubit by admixing a high-lying Rydberg level to the ground-state manifold that couples to a cavity, see Fig.\,\ref{Fig:RydbergDressingScheme}.
%The first part of the scheme, denoted as Dipole 2 in Fig.\,\ref{Fig:RydbergDressingScheme}, facilitates a coupling of the atoms to the light field with both co- and counter rotating terms whose strength is tunable by laser parameters \cite{dimer07}. 

The atomic levels we consider to realize an effective spin system could be the hyperfine-structure manifold of the ground states of ${}^{87}$Rb. Typically this is the $5^2 S_{1/2}$ state split into the $F=1$ and the $F=2$ manifold such that $\ket{0}=\ket{\downarrow}=\ket{F=1,m_F=-1}$ and $\ket{1}=\ket{\uparrow}=\ket{F=2, m_F=-2}$. Here, cavity-assisted Raman transitions couple the $(\ket{0},\ket{1})$ ground-states via adiabatic elimination of the detuned 
excited states ($\ket{d},\ket{e}$) to the cavity
\cite{dimer07,baden14}. Then, the cavity is (indirectly) pumped with photons from the transversal pumping-laser 
that scatter off the atoms and populate the cavity mode. In that way, the pump is 
``hidden'' in the atom-photon coupling $g$: it is the counter-rotating terms that stabilize non-trivial 
steady-states with finite excitation number.

%

%Cavity-assisted Raman transitions were theoretically proposed in 2007 \cite{dimer07} to realise a Dicke superradiance transition in a cavity and were realised experimentally in 2014 \cite{baden14}.
The second part of the pumping scheme, denoted as Dipole 1 in Fig.\,\ref{Fig:RydbergDressingScheme}, consists of admixing a small part of a Rydberg state to the state $\ket{1}$ that is also coupled to the cavity. 
%This is achieved by inducing  (virtual) transitions between Rydberg level and $|1 \rangle$ with a laser of Rabi frequency $\Omega_{\rm Ryd}$ 
%detuned by $\Delta_{\rm Ryd}$ from the Rydberg level. 
To first-order in perturbation theory of the driving, the ground-state becomes dressed with a small fraction of the excited state $\ket{\tilde{1}}\approx \ket{1}+\frac{\Omega_{\rm Ryd}}{2\Delta_{\rm Ryd}}\ket{\rm Ryd}+\mathcal{O}\l\frac{\Omega_{\rm Ryd}}{2\Delta_{\rm Ryd}}\r^2$, where $\Omega_{\rm Ryd}$ is the Rabi-frequency and $\Delta_{\rm Ryd}$ is the detuning from the Rydberg level.
Ground-states $\{\ket{\tilde{1}}_i,\ket{\tilde{1}}_j\}$ interact then with a dressed Rydberg interaction that can be controlled by changing $(\Omega_{\rm Ryd},\Delta_{\rm Ryd})$ of the dressing laser \cite{Pfau10,ZollerP10,Pohl10,GlaetleZoller12}. Dressing schemes for Rydberg atoms on optical lattices have recently been experimentally realised \cite{zeiher16} and the effective Rydberg potential depends strongly on the chosen Rydberg states \cite{glaetzle14}. An additional degree of freedom that allows for engineering anisotropic effective potentials with an angular dependence $V^{\rm eff}_{ij}(r_{ij},\theta_{ij})$ can be introduced by employing states with angular momentum such as states from $P$-manifolds \cite{glaetzle14}.

In a suitably chosen rotating frame of reference, derived in Appendix \ref{sec:DHDR}, the 
parameters appearing in the Hamiltonian Eqs.~(\ref{eq:H_spin},\ref{eq:H_spin-light}), can be 
expressed as follows. For the spin longitudinal field $\Delta$ and the effective 
cavity frequency $\omega_0$, we have
\begin{align}
\Delta =  -\Delta_1+\frac{\Omega^2_{\rm Ryd}}{4\Delta_{\rm Ryd}} \;,\;\;\;\;\;\;
%\nonumber\\
\omega_0 = N \frac{g^2_e}{\Delta_e} +\omega_a \;,
\end{align}
with $\omega_a$ and $\Delta_1$ defined in Eq.\,\eqref{EQ:Rotatingfrequencies}. The coherent coupling 
to the cavity can be tuned by the two external lasers $\Omega_{d,e}$:
\begin{align}
g&=\sqrt{N}\frac{g_{d,e}\Omega_{d,e}}{2\Delta_{d,e}}, \quad \frac{\Omega_d}{4\Delta_d}=\frac{\Omega_e}{4\Delta_e}, \quad \frac{g^2_d}{\Delta_d}= \frac{g^2_e}{\Delta_e}\;.
\label{eq:g}
\end{align}
Finally, the Rydberg-mediated potential takes the general form \cite{Pohl10}
\begin{align}
V^{\rm eff}_{\ell m }&=\l\frac{\Omega_{\rm Ryd}}{2\Delta_{\rm Ryd}}\r^4 \frac{C_6}{r^6_{\ell m }+R^6_c}\;,
\label{EQ:EffRydbergPotential}
\end{align}
and the $V$ appearing in Eq.~(\ref{eq:H_spin}) evaluates Eq.~(\ref{EQ:EffRydbergPotential}) 
at a fixed $r_{\ell m}$ equal to the distance between 
neighboring lattice sites. Effectively step-like potentials 
are also possible. $R_c$ is the critical radius defined by $2 \Delta_{\rm Ryd} \equiv V(R_c)$ which yields $R_c=\l \frac{C_6}{2\hbar |\Delta_{\rm Ryd}|} \r^{1/6}$.
At smaller distances $r_{\ell m} < R_c$, dressing to doubly excited states 
%$\ket{\rm Ryd_{i,j}}$ 
becomes ineffective because of the large detuning $|\Delta_{\rm Ryd}|+V_{\ell m }$. 
The soft-core potentials contain a number of additional resonances 
at $r_{ij} \ll R_c$,
%that result from crossings of Born-Oppenheimer surfaces and 
which are undesirable to realise clean interactions. 
To ensure the interacting atoms in an optical lattice interact via the clean van-der-Waals tail,
it is more advantageous to address relatively low-lying Rydberg states with principal quantum numbers $n\sim 30$,
as $R_c$ can then also shrink down to optical wavelengths. %If the atoms are arranged in an optical lattice with spacings on the order of or bigger than $R_c$, couplings to next-nearest neighbours are weakened as they are in the van der Waals tails $V^{\rm eff}_{ij}\sim \l\frac{\Omega_{\rm Ryd}}{2\Delta_{\rm Ryd}}\r^4 \frac{C_6}{r_{ij}^6}$, since $R_c \sim (C_6)^{1/6}$
% it might be more suitable  since $|C_6|\sim n^{11}$.
Additionally, we comment that complementary to optical lattices, two-dimensional arrays of microtraps have already been used \cite{Browaeys14} to trap single ${}^{87}$Rb atoms with lattice spacings $\sim \mu m$. This would allow to use more highly excited Rydberg states for the dressing interaction bringing with it the advantage of longer lifetimes 
of higher lying Rydberg states.
%for the dressed ground-states $\tau_{\tilde{1}} \sim \tau_{\rm Ryd}/ \l\Omega_{\rm Ryd} /(2\Delta_{\rm Ryd})\r^2 $, where $\tau_{\rm Ryd}$ is the lifetime of the bare Rydberg state.

In App.~\ref{app:blackbodydecays}, we present an overview of the relevant energy and time scales 
including a discussion on problematic Rydberg decays.

%%%%%%%%%%%%%%%%%%%%%%%%%%%%%%%%%%%%%%%%%%%%%%%%%
%\subsection{Mean-field equation of motion}
\section{Coupled even-odd sublattice mean-field master equations for atoms and photons}
\label{sec:langevin}

We now derive and solve the coupled mean-field master equations for both, the spin degrees of freedom and the photons for an infinite number 
of atomic spins $N\rightarrow \infty$.
In absence of the short-range, nearest-neighbor interaction $V$, a mean-field ansatz actually represents the 
exact solution in the long time limit $t\rightarrow \infty$ \cite{gelhausen16}. We account for different spin expectation values on the 
even versus odd sublattice of the bipartite square lattice of Fig.~\ref{Fig:ExperimentalSetUp}. The goal 
is to allow for steady-states with spontaneously broken translational (even-odd interchange) symmetry.

To this end, we now approximate the solution of the full master equation
\begin{align}
\partial_t\rho=-i[H,\rho]+\mathcal{L}_{\kappa}[\rho]+\mathcal{L}_{\gamma}[\rho]\;,
\end{align}
by factorizing the spin part of density operator for all even sites as $\rho_e = \bigotimes_{\ell=1}^{N/2}\rho_{e,\ell}$ 
and analogously for the odd sites $\rho_o = \bigotimes_{\ell=1}^{N/2}\rho_{o,\ell}$.
%Note that this ansatz is {\em exact} in the regime where the photon expectation value $\braket{a}$ either vanishes or is infinitesimally small because the Hamiltonian for $\braket{a}=0$ contains no quantum dynamics and therefore 
%the $\sigma^z_\ell$ are conserved. Therefore the ansatz allows to calculate all 2nd order phase transitions exactly where superradiance sets in.
In the conclusions, Sec.~\ref{sec:conclusions}, we comment further on the prospects of 
capturing finite spatial correlations and fluctuations beyond mean-field.
We further define the spin expectation values on the even and odd sublattices, respectively: 
$\langle \sigma^{\alpha}_{e/o}\rangle=\text{Tr}[\rho_{e/o} \sigma^\alpha_{e/o}]$ 
where $\alpha$ refers to $(x,y,z)$. These six real-valued spin projections are complemented 
by two variables for the photon fields, $\langle a \rangle$, $\langle a^\dagger \rangle $, making it 
a total of eight variables to keep track of. This way, $\rho$ also includes entries to discriminate between the 
vacuum and a coherent photon field.
The first four equations read
\begin{align}
\partial_t\braket{\sigma^x_e(t)}=&\braket{\sigma^y_e(t)} [\Delta-2V( \braket{\sigma^z_{o}(t)} +1 )]-\frac{\gamma}{2}\braket{\sigma^x_e(t)} \label{EQ:Sigmax1}\\
\partial_t\braket{\sigma^y_e(t)}=&\braket{\sigma^x_{e}(t)} [2V(\braket{\sigma^z_{o}(t)} +1)-\Delta ]-2 g  [\braket{a(t)}+\braket{a^{\dagger}(t)}] \braket{\sigma^z_e(t)}-\frac{\gamma}{2}\braket{\sigma^y_e(t)} \\
\partial_t\braket{\sigma^z_e(t)}=&2 g  [\braket{a(t)}+\braket{a^{\dagger}(t)}]\braket{\sigma^y_e(t)}-\gamma(1+\braket{\sigma^z_e(t)}) \label{EQ:Sigmaz1} \\
\partial_t\braket{a(t)}=&-(\kappa +i \omega_0)\braket{a(t)}-\frac{1}{2} i g  (\braket{\sigma^x_{e}(t)}+\braket{\sigma^x_{o}(t)})\label{EQ:PhotonTerm}
\end{align}
The equations for the odd sublattice spin projections follow from Eqs.~(\ref{EQ:Sigmax1}-\ref{EQ:Sigmaz1}) by exchanging 
the sublattice index $e \leftrightarrow o$. The complex conjugate of Eq.~(\ref{EQ:PhotonTerm}) 
completes the set of eight coupled equations. Here, we rescaled the photonic variable 
with $a(t) \to \sqrt{N}\braket{a(t)}$ 
a steady-state is macroscopically occupied in the thermodynamic limit 
and one may also define 
$\braket{\sigma^{\alpha}_{e,o}}\equiv \frac{1}{(N/2)}\sum\limits_{\ell \in even,odd}^N \sigma^{\alpha}_{\ell}$. 

Mean field master equations are often a first step to study driven-dissipative systems, 
see for example, Refs.~\onlinecite{schirmer10,tony11,tony13,tomadin10,tomadin11,everest16,nissen12} and Ref.~\onlinecite{str94} 
for a variety of contexts.
%%%%%%%%%%%%%%%%%%%%%%%%%%%%%%%%%%%%%%%%%%%%%%%%%%%%%%
\section{Derivation and detailed discussion of results}
\label{sec:results}
%

%{\color{blue} Jan4, Sec.~\ref{sec:results} and the Appendices are yours. Pls give it your best, you can also optimize formulations, etc. Possibly it is a good idea, to remove all the text that you have written so far and start from zero again.
%I will try to change/remove as little as possible, when it's done. No double-dipping, underpin 
%the results of the Key results section with "computational meat" (in research paper style not textbook style) 
%and concise explanations of how 
%the results can be obtain - but no repetition or 
%similar formulations.}

\subsection{Result 1: Combination of superradiance and magnetic translation symmetry-breaking}
\label{subsec:result1}
\subsubsection{Phase boundaries and order parameters with photon losses $(\kappa \neq 0, \gamma = 0)$}

We first consider the case where the atoms do not decay spontaneously and the only loss-process is given by the Lindblad $\mathcal{L}_{\kappa}$, see Eq.\,\eqref{EQ:lindbladphoton}.
The Eqs.~(\ref{EQ:Sigmax1}-\ref{EQ:PhotonTerm}) with $\gamma=0$ conserve in this case a pseudo-angular momentum
\begin{align}
\braket{\sigma^x_{e,o}}^2+\braket{\sigma^z_{e,o}}^2=1
\label{EQ:PseudoangularmomentumConstraint}
\end{align}
provided we start from a low-entropy initial state for the spins (as discussed above 
in Subsec.~\ref{intro:result1}), which fulfill this condition.
Here $(e,o)$ refers to the even and odd sub lattice respectively. 
Due to the presence of time-reversal symmetry in the atomic channel, the steady-state of the system constrains $\braket{\sigma^y_{e,o}}=0$ (see the discussion above in Sec.~\ref{intro:result3}). 
%Eqs.~(\ref{EQ:Sigmax1}-\ref{EQ:PhotonTerm}) with $\gamma=0$ can be integrated numerically.
%\begin{figure}[h!]
%\includegraphics[width=11cm]{Integrationofkappaonly.pdf}
%\caption{Numerical integration of the Eqs.~(\ref{EQ:Sigmax1}-\ref{EQ:PhotonTerm}) with $\gamma=0$ with parameters: $V=2.0|\Delta|,\omega_0=2.0|\Delta|,\Delta=-1,\kappa=0.2|\Delta|$. The system eventually reaches a steady-state at long times. The $\braket{\sigma^y_{e,o}}$ magnetisations are protected by symmetry and decay to 0. The relaxation rates to the steady-states can then be calculated from the poles of the inverse-retarded Green function, see \eq{EQ:InverseGreenFunction}.} 
%\label{Fig:ExpSetup}
%\end{figure}
%
Fig.\,\ref{Fig:DRPhaseDiagramkappaonly} is calculated, with photon losses, by numerically solving for the stationary states of Eqs.~(\ref{EQ:Sigmax1}-\ref{EQ:PhotonTerm}). We determine stability by inspecting the real parts of the eigenvalues from the corresponding stability matrix that is obtained by linearising Eqs.~(\ref{EQ:Sigmax1}-\ref{EQ:PhotonTerm}) to first oder in fluctuations around the steady-states.
%The ordered phases are characterised by either a broken $\mathbbm{Z}_2$-symmetry, a broken translational invariance $T_{\rm lat}$ or a phase where both symmetries are spontaneously broken. We observe a phase where all spins are fully polarised (FP) that does not break the aforementioned symmetries. We further identify a plain antiferromagnet (AFM),  a super radiant phase $({\rm SR}_{\rm UNI})$ and most importantly, a coexistence regime (AFM+SR), where the system is characterized by a canted antiferromagnetic excitation pattern on the even and odd sub lattice in addition to a coherent photon condensate, see Tab.\,\ref{Tab:OrderParameters}. 

\begin{figure}
\subfloat[]{\includegraphics[width=9.1cm]{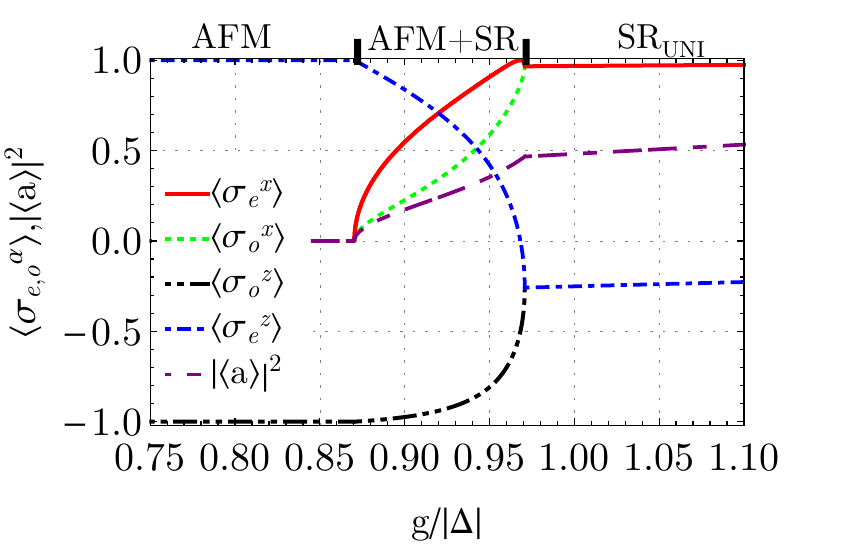}\label{subfig:AFTransitions}}
\subfloat[]{\includegraphics[width=8cm]{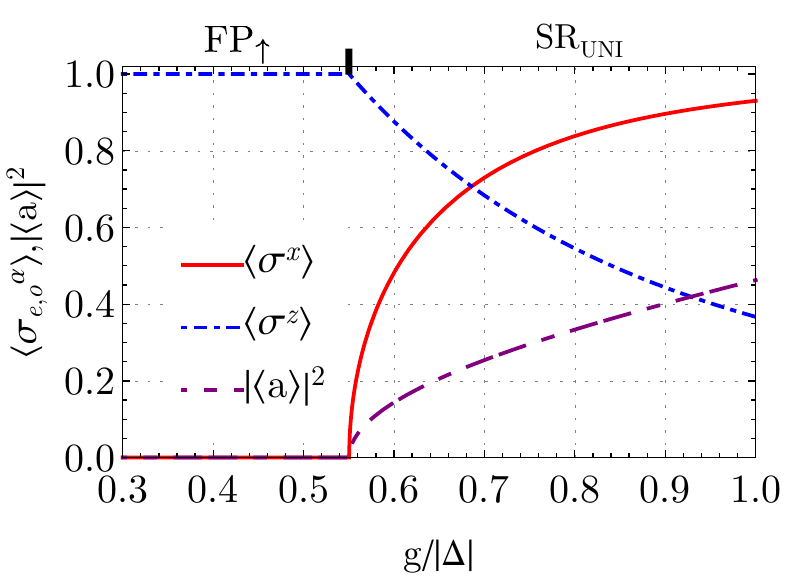}\label{subfig:DickeTransitionshomogeneous}}
\caption{Steady-states for Eqs.~(\ref{EQ:Sigmax1}-\ref{EQ:PhotonTerm}) with $\gamma=0$ for the parameters $\omega_0=2.0|\Delta|, \kappa=0.2|\Delta|$ for the phases depicted in Fig.\,\ref{Fig:DRPhaseDiagramkappaonly}. (a) Behavior of the magnetisations and the coherent photon condensate in the plain antiferromagnet (AFM), in the regime where translational symmetry breaking and a superradiant photon condensate occur together (AFM+SR) and in the Dicke-phase $({\rm SR_{UNI}})$ as the atom-light coupling $g$ is increased at a fixed value of $V=|\Delta|$. All transitions are continuous in the order-parameters. The plot in (b) shows the onset of superradiance as the atom-light coupling is increased, for a fixed value of $V=0.1|\Delta|$.}
\label{Fig:MagneticValues}
\end{figure}

Together with the constraint Eq.\,\eqref{EQ:PseudoangularmomentumConstraint}, the set of  Eqs.~(\ref{EQ:Sigmax1}-\ref{EQ:PhotonTerm}) can be solved analytically.  The homogeneous ($\braket{\sigma^{x,z}}\equiv\braket{\sigma^{x,z}_e}=\braket{\sigma^{x,z}_o}$) steady-state solutions, which describe the Dicke superradiance transition are
\begin{align}
\braket{a}&=\mp\frac{g \braket{\sigma^x}}{\omega_0-i \kappa}\label{EQ:Dickea},\\
\braket{\sigma^x}&=\mp \frac{\sqrt{J-J_c}\sqrt{4J+\Delta}}{\sqrt{2J+V}}\label{EQ:Dickex},\\
\braket{\sigma^z}&=\frac{J_c+\frac{\Delta}{4}}{2J-J_c+\frac{\Delta}{4}}\label{EQ:Dickez}.
\end{align}
We have defined $J=\frac{g^2\omega_0}{\kappa^2+\omega^2_0}$ and $J_c=\frac{\Delta}{4}-V$. 
A plot of the Eqs.\,\eqref{EQ:Dickea}-\eqref{EQ:Dickez} is illustrated in Fig.\,\ref{subfig:DickeTransitionshomogeneous}. Starting in the antiferromagnetic phase and increasing the atom-light coupling $g$, a regime where superradiance $\left(\braket{a}\neq 0 \right)$ and a phase with different sub-lattice magnetisations occur together is predicted $(\braket{\sigma^z_e}-\braket{\sigma^z_o} \neq 0)$ which, due to Eq.\,\eqref{EQ:PseudoangularmomentumConstraint} implies $\braket{\sigma^x_e}-\braket{\sigma^x_o}\neq 0$. Due to the finite longitudinal field $\Delta$, one sublattice is easier flipped than the other and the system realizes a "canted" antiferromagnet. If the atom-light coupling is increased even further, translational symmetry is restored and the system realises a Dicke superradiant phase. This can be seen by tracking the evolution of the magnetisation as $g$ is increased in Fig.\,\ref{subfig:AFTransitions}.

%\begin{figure}[h!]
%\includegraphics[width=8cm]{DiscontinuousMagnetisations.pdf}\label{subfig:Discontinuities}
%\caption{Discontinuous transitions for the steady-state solutions of the magnetizations, corresponding to the dotted arrow in Fig.\,\ref{Fig:DRPhaseDiagramkappaonly}. Plot shows the magnetizations for $\omega_0=2.0, \kappa=0.2, -\Delta=-1$ and $g=g_{\rm c,1}(V=0)$, see \eq{EQ:gcrit1}. Here we have labelled $\braket{\sigma^{x,z}}\equiv\braket{\sigma^{x,z}_{e}}=\braket{\sigma^{x,z}_{o}}$. }
%\label{Fig:Discontinuities}
%\end{figure}
%%%%%%%%%%%%%%%%%%%%%%%%%%%%%%%%%

%%%%%%%%%%%%%%%%%%%%%%%%%%%%%%%%%%%%%
We now derive analytical expressions for the phase-transition lines.
%The Heisenberg-Langevin Equations are given by 
%\begin{align}
%\partial_t\sigma^x_e(t)=&\sigma^y_e(t) [\Delta-2V( \sigma^z_{o}(t) +1 )]-\frac{\gamma}{2}\sigma^x_e(t)+\mathcal{F}_{x,e}(t)\label{EQ:HeisenbergLangevinx}\\
%\partial_t\sigma^y_e(t)=&\sigma^x_{e}(t) [2V(\sigma^z_{o}(t) +1)-\Delta ]-2 g  [a(t)+a^{\dagger}(t)] \sigma^z_e(t)-\frac{\gamma}{2}\sigma^y_e(t)+\mathcal{F}_{y,e}(t) \\
%\partial_t\sigma^z_e(t)=&2 g  [a(t)+a^{\dagger}(t)]\sigma^y_e(t)-\gamma(1+\sigma^z_e(t))+\mathcal{F}_{z,e}(t) \\
%\partial_t\braket{a(t)}=&-(\kappa +i \omega_0)a(t)-\frac{1}{2} i g  (\sigma^x_{e}(t)+\sigma^x_{o}(t))+\mathcal{F}_{a}(t)\label{EQ:HeisenbergLangevina}
%\end{align}
First, we transform the mean-field equations (\ref{EQ:Sigmax1}-\ref{EQ:PhotonTerm}) in frequency space via Fourier transformation
\begin{align}
\mathcal{O}(t)=\frac{1}{\sqrt{2\pi}}\int\limits_{-\infty}^{\infty}e^{-i\nu t}\mathcal{O}(\nu) d\nu, \quad \mathcal{O}^{\dagger}(t)=\frac{1}{\sqrt{2\pi}}\int\limits_{-\infty}^{\infty}e^{-i \nu t}\mathcal{O}^{\dagger}(-\nu) d\nu
\label{EQ:Fourier-Transformation}
\end{align}
In general, one should add Markovian quantum noise-operators with zero-mean to the photonic and atomic set of the mean-field master equations that result from the interaction of the atom-cavity system with the vacuum modes outside of the cavity. These we denote $\mathcal{F}^{\alpha}_{e,o}(\nu)$ as the atomic and $\mathcal{F}^{a}(\nu)$ as the photonic noise-operators in frequency space. 

Next, we add back fluctuations to Eqs.~(\ref{EQ:Sigmax1}-\ref{EQ:PhotonTerm})
\begin{align}
\braket{\sigma^{\alpha}_{e,o}(\nu)}&\to \braket{\sigma^{\alpha}_{e,o}}\delta(\nu)\sqrt{2\pi}+\delta\sigma^\alpha_{e,o}(\nu)\\
\sqrt{N}\braket{a(\nu)}&\to \sqrt{N}\braket{a}\delta(\nu)\sqrt{2\pi}+\delta a(\nu)\;,
\end{align}
where the steady-states are denoted as $\braket{\sigma^{\alpha}_{e,o}}$ with $\alpha=(x,y,z)$ and $\braket{a}$ is the expectation value for a coherent photon condensate.  Here, $\delta \sigma^{\alpha}_{e,o}(\nu)$ and $\delta a(\nu)$ describe quantum fluctuations about the semi-classical steady-state and $\delta(\nu)$ denotes a delta function in frequency space. At long times, we may neglect second-order terms in the fluctuations by assuming that the steady-state values are large compared to the associated fluctuations in the thermodynamic limit $N\to \infty$.

The now linearized equations can be cast in matrix form.
\begin{align}
\b{\mathcal{F}}(\nu)=\delta(\nu)f(\b{\sigma})+\b{G}^{-1}_{R}(\nu)\cdot \b{\delta \sigma}(\nu) \label{EQ:LinearLangevinEquation}
\end{align}
where the fluctuations around the steady state are collected in $ \b{\delta \sigma}(\nu)$ and the noise-operators are collected into $\b{\mathcal{F}}(\nu)$:
\begin{align}
\delta\b{\sigma}^{T}(\nu)&=\l \delta \sigma^{x}_e(\nu), \delta \sigma^{y}_e(\nu), \delta \sigma^z_e(\nu),\delta a(\nu), \delta a^{\dagger}(-\nu),\delta \sigma^{x}_o(\nu), \delta \sigma^{y}_o(\nu), \delta \sigma^z_o(\nu) \r \\
\b{\mathcal{F}}^{T}(\nu)&=\l \mathcal{F}^{x}_e(\nu), \mathcal{F}^{y}_e(\nu), \mathcal{F}^{z}_e(\nu), \mathcal{F}^{a}(\nu), \mathcal{F}^{a^{\dagger}}(-\nu),\mathcal{F}^{x}_o(\nu),\mathcal{F}^{y}_o(\nu), \mathcal{F}^{z}_o(\nu) \r
\end{align}
The function $f(\b{\sigma})$ is associated with the coherent part of the steady-states and thus only leads to a zero-frequency peak in the cavity-spectrum. The responses of the fluctuations $\delta\b{\sigma}$ to the noise or 'driving-forces' $\b{\mathcal{F}}$ is encoded by the retarded Green-function $\b{G}_{R}(\nu)$, its inverse $\b{G}^{-1}_{R}(\nu)$ is given as:
\begin{align}
\left(\scalemath{0.75}{
\begin{array}{cccccccc}
 \frac{1}{2} (\gamma -2 i \nu ) & 2 V (\braket{\sigma^z_o}+1)-\Delta  & 0 & 0 & 0 & 0 & 0 & 2 V \braket{\sigma^y_e} \\
 \Delta -2 V (\braket{\sigma^z_o}+1) & \frac{1}{2} (\gamma -2 i \nu ) & 2 g (\braket{a} +\braket{a^{\dagger}}) & 2 g \braket{\sigma^z_e} & 2 g \braket{\sigma^z_e} & 0 & 0 & -2 V \braket{\sigma^x_e} \\
 0 & -2 g (\braket{a} +\braket{a^{\dagger}}) & \gamma -i \nu  & -2 g \braket{\sigma^y_e} & -2 g \braket{\sigma^y_e} & 0 & 0 & 0 \\
 \frac{1}{2}i g & 0 & 0 & \kappa -i (\nu -\omega_0) & 0 & \frac{1}{2} i g & 0 & 0 \\
 -\frac{1}{2} i g & 0 & 0 & 0 & \kappa -i (\nu +\omega_0) & -\frac{1}{2} i g & 0 & 0 \\
 0 & 0 & 2 V \braket{\sigma^y_o} & 0 & 0 & \frac{1}{2} (\gamma -2 i \nu ) & 2 V (\braket{\sigma^z_e}+1)-\Delta  & 0 \\
 0 & 0 & -2 V \braket{\sigma^x_o} & 2 g \braket{\sigma^z_o} & 2 g \braket{\sigma^z_o} & \Delta -2 V (\braket{\sigma^z_e}+1) & \frac{1}{2} (\gamma -2 i \nu ) & 2 g (\braket{a} +\braket{a^{\dagger}}) \\
 0 & 0 & 0 & -2 g \braket{\sigma^y_o} & -2 g \braket{\sigma^y_o} & 0 & -2 g (\braket{a} +\braket{a^{\dagger}}) & \gamma -i \nu  \\
\end{array}}
\right)
\label{EQ:InverseGreenFunction}
\end{align}
The frequency-resolved spectrum of excitations governed by the fluctuations can be obtained from the characteristic equation 
\begin{align}
{\rm Det}[\b{G}^{-1}_{\rm R}(\nu)]=0.
\label{EQ:CharacteristicEquationGreenFunction}
\end{align}
All poles of the retarded Green function are located in the lower complex frequency plane. The damping rate of the excitations can be read off from the imaginary part of these poles, see for instance Reference \cite{kamenev_book}.
In the case of second order transitions, the order-parameters $(\braket{a},\braket{\sigma^x_{e,o}},\braket{\sigma^z_{e,o}})$ change continuously at the phase transitions. We obtain analytical expressions for the phase boundaries by solving
\begin{equation}
\lim\limits_{\nu \to 0}{\rm Det}[\b{G}^{-1}_{\rm R}(\nu)]=\alpha^2=0\;.
\end{equation}
The zeroth-order frequency component refers to a possible gap $\alpha^2$ of the system that will close continuously $(\lim_{g\to g_c}\alpha^2\to 0)$ when the phase transition is approached by increasing the atom-light coupling $g$. 
We arrive at the set of transition lines given by \eq{EQ:Vcrit1}, \eq{EQ:Vcrit2} and \eq{EQ:Vcrit3} that are depicted as black lines in Fig.\,\ref{Fig:DRPhaseDiagramkappaonly} where they match the numerically calculated transitions. The open character of the system becomes manifest in the expressions for the phase boundaries as all transitions explicitly depend on the rate of photonic dissipation $\kappa$. Starting from the FP$_{\uparrow}$ phase, the Dicke superradiance transition in presence of the Rydberg interaction sets in at the critical coupling strength:
\begin{align}
g_{\rm c,1}&=\frac{\sqrt{\kappa ^2+\omega_0^2} \sqrt{\omega_0 (\Delta -4 V)}}{2 \omega_0},\label{EQ:gcrit1}\\
V_{\rm c,1}&=\frac{\Delta }{4}-\frac{g^2 \omega_0}{\kappa ^2+\omega_0^2}.\label{EQ:Vcrit1}
\end{align}
The finite Rydberg-dressed interaction $V$ modifies the effective longitudinal field experienced by the spins which shifts the position of the superradiant condensate in comparison to the $V=0$ case. Eq.\,\eqref{EQ:gcrit1} collapses to the familiar Dicke superradiance transition in the case $V\to 0$ \cite{dimer07}. 
The crossover from the (AFM+SR) regime to the Dicke superradiant phase (SR) is given by:
\begin{align}
&V_{\rm c,2}=\nonumber\\
&\bigg[4 g ^2 \omega_0 \left(4 g ^2 \omega_0+\sqrt{\Delta ^2 \left(\kappa ^2+\omega_0^2\right)^2-8 \Delta  g ^2 \omega_0 \left(\kappa ^2+\omega_0^2\right)+80 g ^4 \omega_0^2}\right)\nonumber\\
&+\Delta  \left(\kappa ^2+\omega_0^2\right) \sqrt{\Delta ^2 \left(\kappa ^2+\omega_0^2\right)^2-8 \Delta  g ^2 \omega_0 \left(\kappa ^2+\omega_0^2\right)+80 g ^4 \omega_0^2}\nonumber\\
&+\Delta ^2 \left(\kappa ^2+\omega_0^2\right)^2\bigg]/8 \left(\kappa ^2+\omega_0^2\right) \left(\Delta  \left(\kappa ^2+\omega_0^2\right)+2 g ^2 \omega_0\right)\;.
\label{EQ:Vcrit2}
\end{align}
The transition line from the AFM phase to the AFM+SR regime is given by $(-\Delta<0)$
\begin{align}
g_{\rm c,3}&=\frac{\sqrt{\Delta } \sqrt{\kappa ^2+\omega_0^2} \sqrt{4 V-\Delta }}{2 \sqrt{2} \sqrt{V} \sqrt{\omega_0}},\\
V_{\rm c,3}&=\frac{\Delta ^2 \left(\kappa ^2+\omega_0^2\right)}{4 \left(\Delta  \left(\kappa ^2+\omega_0^2\right)-2 g ^2 \omega_0\right)}\;.
\label{EQ:Vcrit3}
\end{align}
We note that the line where AFM and SR order occur together diverges $V_{\rm c,3} \to \infty$ as $\lim_{g \to g_{\star}}$ with
$g_{\star}=\l\sqrt{\Delta } \sqrt{\kappa ^2+\omega_0^2}\r/\l\sqrt{2} \sqrt{\omega_0}\r$
Moreover, on a mean-field level, there is a touching point $g_t$ of two second-order phase transition lines that can be found by equating \eq{EQ:Vcrit2} and \eq{EQ:Vcrit3} which yields
$g_t=\l\sqrt{\Delta}\sqrt{\kappa^2+\omega^2_o}\r\l2\sqrt{\omega_o}\r$ and $  V_{\rm c,3}(g_t)=\Delta/2$ marks the point where the effective longitudinal field on one of the sublattices vanishes.
%\subsection{Fluctuations}
%In the pure Dicke-phase, the Hamiltonian corresponds to that of a classical Ising model coupled to a transverse field. The interaction of the cavity photons with the light-field mediates an infinite-range atom-atom interaction in $x$-direction. Note that the ferromagnetic coupling constant is dissipative in a non-equilibrium framework and reflects photon loss through the cavity that affect the atom-light coupling strength.
%\begin{align}
%H_{\rm Dicke}=J \sum\limits_{\ell m } \sigma^x_m \sigma^x_{\ell}+\frac{\Delta}{2}\sum_{\ell=1}^N \sigma^z_{\ell}
%\end{align} 
%where the coupling constant is given by
%\begin{align}
%J= \frac{g^2}{N}\frac{\omega_0}{\omega_0^2+\kappa^2}
%\end{align}
%this can be seen for example by solving for \Eq{EQ:PhotonDissipation} and re-inserting into the other equations. Deviations from the dynamics of the transverse Ising model are suppressed by $1/N$-corrections. 
%In the plain antiferromagnetic phase, the Hamiltonian corresponds to an anti-ferromagnet in $z$-direction coupled to a longitudinal field. Inter-site fluctuations .  
%\begin{align}
%H_{\rm AF}=\sum_{\braket{\ell m}}^{N}\sigma^{ee}_{\ell}\sigma^{ee}_m-\frac{\Delta}{2}\sum_{\ell=1}^N\sigma^z_{\ell}\;,
%\end{align}
On a mean-field level, we find a multi-critical point, where all second-order phase transition lines meet on the $g=0$-axis at $V=\Delta/4$.
%Fluctuations will likely modify this finding and shift the phase boundaries. 

In Appendix \ref{Sec:Equilibrium}, we analyze a $T=0$ equilibrium spin model with the same 
phases as Fig.\,\ref{Fig:DRPhaseDiagramkappaonly} upon identifying one spin interaction constant 
with cavity parameters. Dynamics and statistics remains drastically different in the 
non-equilibrium case, however.

%%%%%%%%%%%%%%%%%%%%%%%%%%%%%%%%%%%%%%%%%%%%

\subsection{Result 2: Even-odd sublattice peak in cavity spectrum}
\label{subsec:result2}
\subsubsection{Derivation of the cavity output spectrum ($\kappa \neq 0, \gamma=0$)}
\begin{figure}[t!]
\subfloat[]{\put(108,155){AFM-phase}\includegraphics[width=8cm]{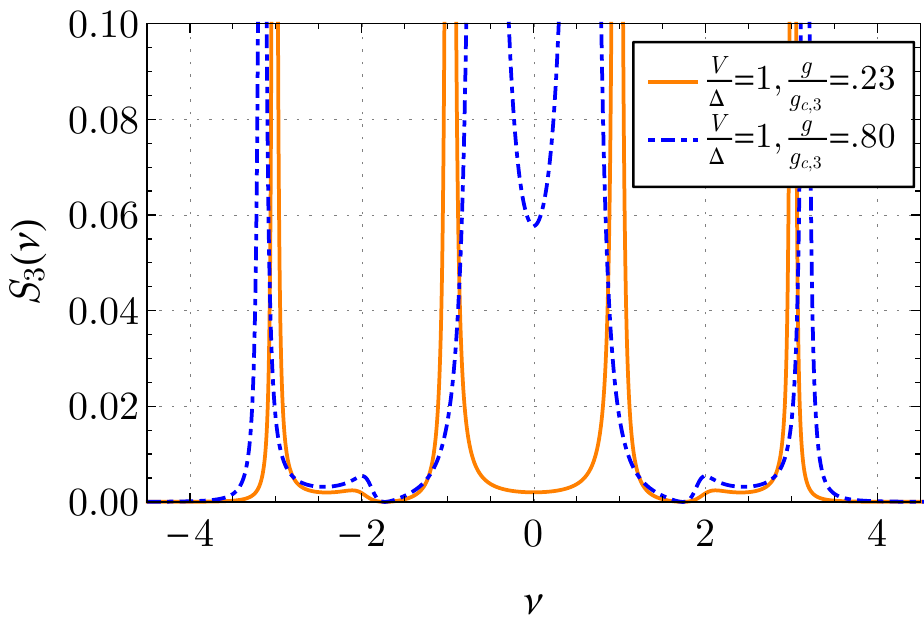}\label{subfig:CavSpectrTrivAF}}
\subfloat[]{\put(108,155){AFM+SR-phase}\includegraphics[width=8cm]{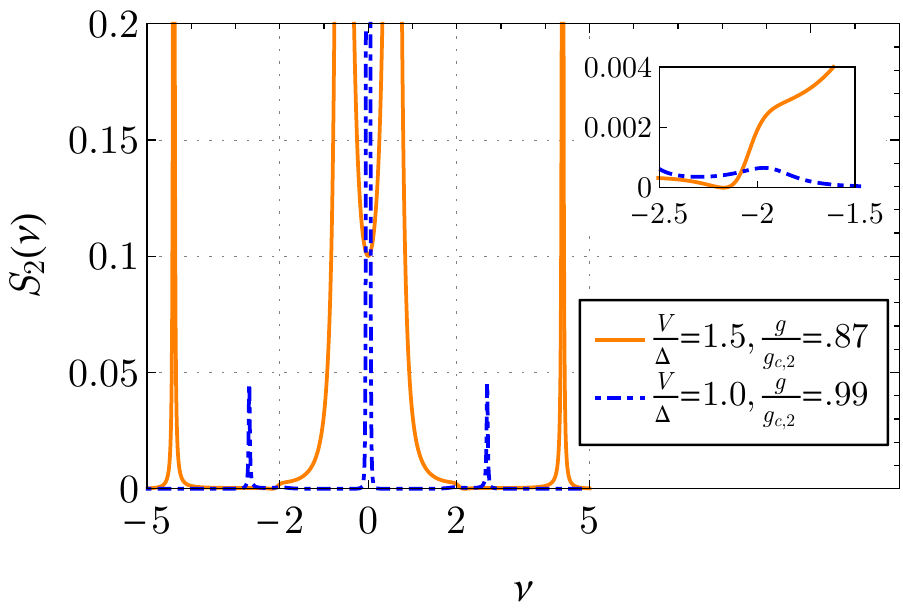}\label{subfig:DickeRydbergTransitions}}\\
\subfloat[]{\put(108,155){${\rm FP}_{\uparrow}$-phase}\includegraphics[width=8cm]{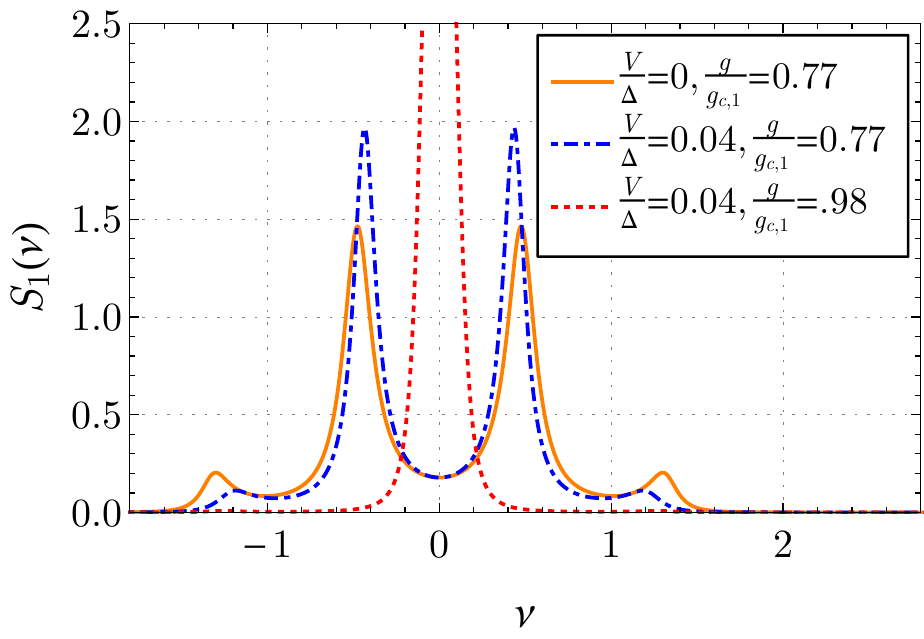}\label{subfig:UnPolKappaTransitions}}
\subfloat[]{\put(108,155){${\rm SR}_{\rm UNI}$-phase}\includegraphics[width=8cm]{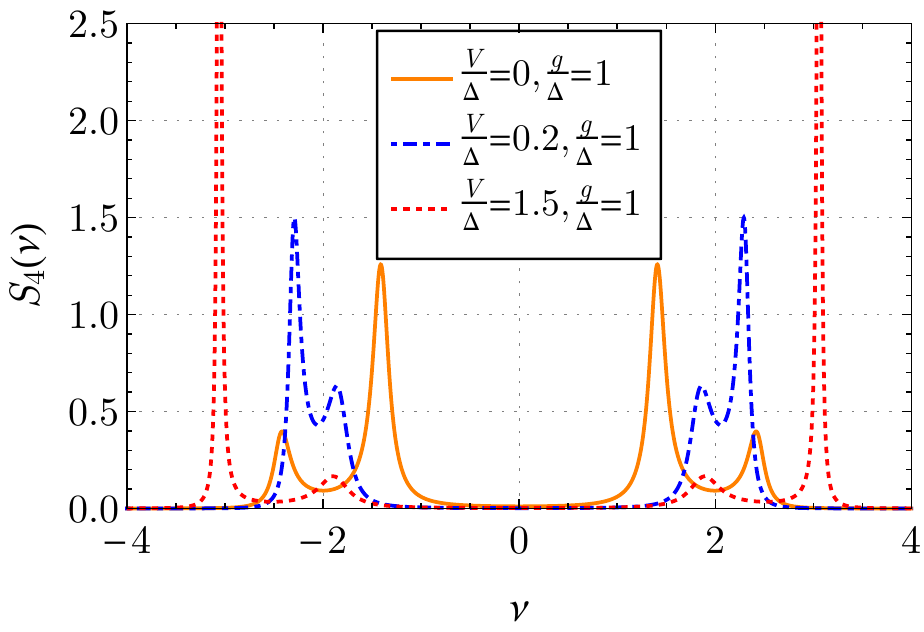}\label{subfig:DickeTransitions}}
\caption{Typical cavity spectra for each of the four phases depicted in Fig.\,{\ref{Fig:DRPhaseDiagramkappaonly}}. In general, the resonances show the hybridised atom-cavity eigenenergies that can be obtained from solving Eq.\,\eqref{EQ:CharacteristicEquationGreenFunction}. There are four poles in the cavity spectrum when translational symmetry is intact as in (c,d) and there are six poles when translational symmetry is broken as (a,b).
(a) Cavity spectrum in the plain antiferromagnetic phase. The broken translational symmetry is reflected in the appearance of an additional (Rydberg) resonance in the atomic-sector. (b) Cavity spectra in the regime of a broken $\mathbbm{Z}_2$ and translational symmetry $T$. As $g\to g_{c,2}$, two of the six poles move towards $\nu=0$ (dot-dashed line). At $g=g_{c,2}$ translational symmetry is restored and the additional Rydberg-induced even-odd peaks disappear. (Inset) Close to the frequency at $\nu=\omega_0/\Delta=\pm 2.0$ there are resonances with small but finite weight corresponding to the cavity resonance. (c) Cavity spectrum for the fully polarized phase FP$_{\uparrow}$ with no photonic excitations $\braket{a}=0$. (d) Spectrum in the superradiant regime $\braket{a} \neq 0$ with translational symmetry $T$ still intact.}
\label{Fig:Cavity Spectra}
\end{figure}

Here we calculate the frequency-resolved cavity output spectrum for the light that leaks from the imperfect cavity mirrors within a standard input-output theory \cite{collett84,gardiner85,dimer07}. We find that every phase in Fig.\,\ref{Fig:DRPhaseDiagramkappaonly} shows a characteristic cavity output spectrum making it possible to experimentally distinguish one phase from the other.
The input fields are related to the output fields by the relation
\begin{align}
a_{\rm out}(\nu)&=\sqrt{2 \kappa}a(\nu)-a_{\rm in}(\nu),\label{EQ:InputOutputRelation}\\
a_{\rm out}^{\dagger}(-\nu)&=\sqrt{2 \kappa}a^{\dagger}(-\nu)-a_{\rm in}^{\dagger}(-\nu).
\end{align}
The annihilation operators $\left(a_{\rm out}(\nu), a_{\rm in}(\nu), a(\nu)\right)$ correspond to the output field, the vacuum input field, and the intra cavity field, respectively and we have used $\mathcal{F}_{a}(\nu)=\sqrt{2\kappa}a_{\rm in}(\nu)$.
%Output and input fields are related by the boundary condition:
%\begin{align}
%a_{\rm out}(\nu)=\sqrt{2\kappa}a(\omega)-a_{\rm in}(\nu)
%\label{EQ:InputOutPutBoundary}
%\end{align}
The Markovian quantum noise operators with zero mean are determined by their second-order correlation functions. For the photonic channel they read
\begin{align}
\braket{a_{\rm in}(\nu')a^{\dagger}_{\rm in}(-\nu)}&=\delta(\nu+\nu')\label{EQ:CorrelationFunction}
\end{align}
%and the correlations of the noise-operators for the atomic channel in a markovian framework are
%\begin{align}
%\braket{\mathcal{F}^i_{\ell}(t)\mathcal{F}^j_{m}(t')}=\gamma M^{ij}\delta_{\ell m}\delta(t-t')=\gamma
%\left(
%\begin{array}{ccc}
% 1 & -i & 2\braket{\sigma^-} \\
% i &  1  & 2i \braket{\sigma^-} \\
% 2\braket{\sigma^+} & -2i\braket{\sigma^+} & 2(1+\braket{\sigma^z})  \\
%\end{array}
%\right)_{ij}\delta(t-t')\delta_{\ell m}
%\label{EQ:NoiseCorrelator}
%\end{align}
%The correlation functions are generic for a markovian bath kept at zero temperature and we refer to \cite{scully} and for a specific derivation to the appendices in \cite{gelhausen16,Marcuzzi16}.
We solve \eq{EQ:LinearLangevinEquation} for $a_{\rm out}(\nu)$ and $a^{\dagger}_{\rm out}(-\nu)$ together with Eq.\,\eqref{EQ:CorrelationFunction} to obtain the output spectrum for a vacuum input field
\begin{align}
S(\nu)=\braket{a^{\dagger}_{\rm out}(\nu)a_{\rm out}(\nu)}
=2\kappa\braket{\delta a^{\dagger}(\nu)\delta a(\nu)}=2\kappa\int_{-\infty}^{\infty}e^{-i\nu \tau}\braket{\delta a^{\dagger}(0)\delta a(\tau)}d\tau.
\label{Eq:CavSpectrumRelation}
\end{align}
%Where we have expressed the cavity spectrum as the steady-state autocorrelation function for the intra-cavity field
%where the subscript $ss$ refers to the limit $\braket{a^{\dagger}(\tau)a(0)}_{\rm ss}=\lim\limits_{t \to \infty}\braket{a^{\dagger}(t+\tau)a(t)}$. The spectrum will consist of a coherent part $S_{\rm coh}(\nu)$ that stems from the steady-states $\left(\braket{a},\braket{\sigma^{\alpha}_{e,o}}\right)$ and an incoherent part that governs the fluctuations $\delta a(\nu) \delta a^{\dagger}(\nu')$. 
The unnormalized fluorescence spectrum $S(\nu)$ is proportional to finding a photon at frequency $\nu$ and thus displays the position and the spectral weight of the resonance energies of hybridized atom-cavity modes. We only depict cavity-spectra in the $\gamma=0$ limit. We have investigated the effect of spontaneous emission on the cavity spectra in the $V\to 0$ limit previously \cite{gelhausen16} and found that it can induce a frequency asymmetry in the cavity spectrum since atomic excitations can leave the cavity directly by emission into free space. The cavity spectra for $\kappa \neq 0, \gamma=0$ can be obtained for every phase in Fig.\,\ref{Fig:DRPhaseDiagramkappaonly}.
In the fully polarized phase ($\braket{a}=\braket{\sigma^x}=0,\braket{\sigma^z}=1$) for $g<g_{c,1}$ and $-\Delta<0$, the cavity spectrum is obtained as 
\begin{align}
S_{1}(\nu)&=\frac{16 \kappa ^2 g ^4 (\Delta -4 V)^2}{|\Omega_0|^2},\\
|\Omega_1|^2&=\left|\left((\kappa -i \nu )^2 (-\Delta +\nu +4 V) (\Delta +\nu -4 V)+4 g ^2 \omega_0 (\Delta -4 V)+\omega_0^2 (-\Delta +\nu +4 V) (\Delta +\nu -4 V)\right)\right|^2\nonumber
\end{align}
and is depicted in Fig.\,\ref{subfig:UnPolKappaTransitions}.
In the limit $V \to 0$ it reduces to the familiar expression obtained in Ref.\,\cite{dimer07}.
In the AFM phase ($\braket{\sigma^z_e}=-1,\braket{\sigma^z_o}=1, \braket{a}=\braket{\sigma^x}=0$), the spectrum is given by
\begin{align}
S_{3}(\nu)=&\frac{64 \kappa ^2 g ^4 V^2 \left(\Delta ^2-4 \Delta  V+\omega ^2\right)^2}{|\Omega_3|^2}\\
\Omega_3=&\left|8 g ^2 V \omega_0 \left(\Delta ^2-4 \Delta  V+\omega ^2\right)+(\omega -\Delta ) (\Delta +\omega ) (\kappa -i \omega )^2 (-\Delta +4 V+\omega ) (\Delta -4 V+\omega )\nonumber\right .\\
&\left. +\omega_0^2 (\omega -\Delta ) (\Delta +\omega ) (-\Delta +4 V+\omega ) (\Delta -4 V+\omega )\right|^2
\end{align}
and a typical spectrum can be seen in Fig.\,\ref{subfig:CavSpectrTrivAF}.
In the homogeneous phase we make use of Eqs.\,(\ref{EQ:Dickea}-\ref{EQ:Dickez}) and obtain the spectrum as 
\begin{align}
S_4(\nu)&=\frac{16 \kappa ^2 g ^8 \omega_0^2 \left(\kappa ^2+\omega_0^2\right)^6 (\Delta -2 V)^4}{|\Omega_4|^2}
\end{align}
with the abbreviated expressions
\begin{align}
\Omega_4&=\left|4 g ^4 \omega_0^2\left(\Omega_{11}+\Omega_{42}\right)+V^2 \left(\kappa ^2+\omega_0^2\right)^2\left(\Omega_{43}+\Omega_{44}+\Omega_{45}\right)+2 g ^2 V \omega_0 \left(\kappa ^2+\omega_0^2\right)\left(\Omega_{46}+\Omega_{47}+\Omega_{48}\right)\right|^2,\nonumber\\
\Omega_{41}&=\Delta ^2 \left(\kappa ^2+\omega_0^2\right)^3+\kappa ^6 \omega ^2+2 i \kappa ^5 \omega ^3-\kappa ^4 \left(\omega ^4-3 \omega ^2 \omega_0^2\right)+4 i \kappa ^3 \omega ^3 \omega_0^2+\omega_0^2 \left(\omega_0^2-\omega ^2\right) \left(\omega ^2 \omega_0^2-16 g ^4\right),\nonumber\\
\Omega_{42}&=\kappa ^2 \omega_0^2 \left(-16 g ^4-2 \omega ^4+3 \omega ^2 \omega_0^2\right)-2 i \kappa  \left(16 g ^4 \omega  \omega_0^2-\omega ^3 \omega_0^4\right),\nonumber\\
\Omega_{43}&=-\kappa ^4 \left(8 \Delta  g ^2 \omega_0+\omega ^4-3 \omega ^2 \omega_0^2\right)+4 i \kappa ^3 \omega  \omega_0 \left(\omega ^2 \omega_0-4 \Delta  g ^2\right)+2 i \kappa  \omega  \omega_0^2 \left(-8 \Delta  g ^2 \omega_0-48 g ^4+\omega ^2 \omega_0^2\right),\nonumber\\
\Omega_{44}&=\kappa ^6 \omega ^2+2 i \kappa ^5 \omega ^3+\kappa ^2 \omega_0 \left(8 \Delta  g ^2 \left(\omega ^2-2 \omega_0^2\right)-32 g ^4 \omega_0-2 \omega ^4 \omega_0+3 \omega ^2 \omega_0^3\right),\nonumber\\
\Omega_{45}&=\omega_0^2 \left(8 \Delta  g ^2 \omega_0 \left(\omega ^2-\omega_0^2\right)+16 g ^4 \left(3 \omega ^2-2 \omega_0^2\right)-\omega ^4 \omega_0^2+\omega ^2 \omega_0^4\right),\nonumber\\
\Omega_{46}&=-8 \Delta  g ^2 \omega_0 \left(\kappa ^2+\omega_0^2\right)^2+\Delta ^2 \left(\kappa ^2+\omega_0^2\right)^2 \left(\kappa ^2+2 i \kappa  \omega -\omega ^2+\omega_0^2\right),\nonumber\\
\Omega_{47}&=2\left(\kappa ^6 \omega ^2+2 i \kappa ^5 \omega ^3-\kappa ^4 \left(\omega ^4-3 \omega ^2 \omega_0^2\right)+4 i \kappa ^3 \omega ^3 \omega_0^2+\omega_0^2 \left(\omega_0^2-\omega ^2\right) \left(\omega ^2 \omega_0^2-24 g ^4\right)\right),\nonumber\\
\Omega_{48}&=2\left(\kappa ^2 \omega_0^2 \left(-24 g ^4-2 \omega ^4+3 \omega ^2 \omega_0^2\right)-2 i \kappa  \left(24 g ^4 \omega  \omega_0^2-\omega ^3 \omega_0^4\right)\right), \nonumber
\end{align}
depicted in Fig.\,\ref{subfig:DickeTransitions}.
We denote the cavity spectrum in the AFM+SR regime as $S_4(\nu)$. We solve Eqs.~(\ref{EQ:Sigmax1}-\ref{EQ:PhotonTerm}) numerically in the long-time limit and use \eq{Eq:CavSpectrumRelation} to determine the spectrum numerically, see Fig.\,\ref{subfig:DickeRydbergTransitions}.
Now we discuss the characteristic features and the behavior of the poles. 
%%%%%%%%%%%%%%%%%%%%%%%%%%%%%%%%%%%%%%%%%%%%%%%%%%%%%%

\subsubsection{Discussion of cavity spectra and low-frequency pole structure for ($\kappa \neq 0, \gamma = 0 $)}
We begin our discussion with an analysis of the cavity spectra $S(\nu)$ in each phase. The cavity spectra are shown in Fig.\,\ref{Fig:Cavity Spectra}.
In general there are either four or six poles in the cavity spectrum. In the former case these originate from two photon-branches and two atomic branches that are symmetrically arranged around the zero-frequency axis. We identify the branches by their $g\to 0$ limit in the fully polarised phase where the resonances settle at the bare frequencies given by $\nu_{\rm Atom}=\pm\Delta$ and $\nu_{\rm Photon}=\pm\omega_{0}-i\kappa$. There are six poles when the translational symmetry in the atomic sector is broken. The additional poles reflect the even/odd imbalance of the system and are thus attributed to the Rydberg interaction, see Fig.\,\ref{subfig:CavSpectrTrivAF}. This provides a clear feature to experimentally detect a phase with antiferromagnetic order. 

We describe and depict the characteristic features of the cavity spectra for each phase of the mean-field phase diagram in Fig.\,\ref{Fig:DRPhaseDiagramkappaonly} below.
The cavity spectra in Fig.\,\ref{subfig:UnPolKappaTransitions} (fully polarized FP$_{\uparrow}$) and in Fig.\,\ref{subfig:DickeTransitions} (superradiant phase) are well-known and derived in \cite{dimer07} in the $V\to 0$ limit. In the superradiant regime, an increasing Rydberg interaction $V$ shifts the atomic poles to higher energies whereas the peaks associated to the photonic branch settle around the cavity resonance at ${\rm Re}(\nu) =\pm \omega_0=\pm2.0|\Delta|$. In the AFM and the (AFM+SR) phase, the cavity spectra depicted in Fig.\,\ref{subfig:CavSpectrTrivAF} and Fig.\,\ref{subfig:DickeRydbergTransitions} exhibit the aforementioned even/odd sublattice peak that reflects the broken translational symmetry in the atomic sector, so that there are six poles in total.

The frequency-resolved eigenenergies of the hybridized atom-cavity modes display a characteristic behaviour close to the phase transition as $g\to g_{\rm c,1,2,3}$ in Fig.\,\ref{Fig:DRPhaseDiagramkappaonly}. On the real frequency axis, all phase transitions appear when at least one of the either four or six poles hits the zero $\nu= 0$.  The low-frequency behaviour of the critical poles leading to the Dicke superradiance transition has already been established \citep{DallaTorre}. From the four poles, two (which we refer to as ($\nu_a, \nu_b$) in the following) approach the origin in the complex frequency plane. First, both poles become completely imaginary as $g\to g_{c,1}$. A single one of these poles vanishes at the phase-transition $|\nu_a|=0$ while the other retains a finite imaginary part at $g=g_{c,1}$ set by the dissipation ($\nu_b(g=g_{c,1}) \sim -i \kappa$) emphasizing that the Dicke superradiance transition directly couples to the dissipation. The cavity spectrum in Fig.\,\ref{subfig:UnPolKappaTransitions} exhibits a pole at zero frequency but with a finite, purely imaginary contribution. The intensity under this finite-width peak diverges, which indicates a macroscopic occupation of the cavity-mode $\braket{a}\neq 0$. As the transition from the AFM into the (AFM+SR)-phase involves a superradiance transition, the same behaviour of the low-frequency poles can be observed as $g \to g_{\rm c,3}$. 
When translational symmetry is broken, this transition mainly affects the atomic channel. 
As the photonic sector alone couples to a dissipative channel, we numerically observe that when translational symmetry is restored as one goes from the (AFM+SR) into the SR phase as $g\to g_{\rm c,2}$, two of the six poles approach the origin on the complex frequency plane. In contrast to the Dicke superradiance transition both the real and imaginary part of the two low-frequency poles vanish together, see Fig.\,\ref{subfig:DickeRydbergPoles} for an illustration of the pole structure and Fig.\,\ref{subfig:DickeRydbergTransitions} for the cavity spectra in the (AFM+SR) phase. At $g>g_{\rm c,2}$ translational symmetry $T_{\rm lat}$ is restored and the spectrum is given by Fig.\,\ref{subfig:DickeTransitions}.

%%%%%%%%%%%%%%%%%%%%%%%%%%%%%%%%%%%%%%%%%%%%%%%%%%%%%%

\subsection{Result 3: Photon number oscillations}
\label{subsec:result3}
\subsubsection{Phase boundaries and order parameters with spontaneous emission 
($\kappa \neq 0, \gamma \neq 0$)}

As discussed in Sec.\,\ref{intro:result3}, allowing for spontaneous emission ($\gamma\neq 0$) in addition to photon leakage ($\kappa \neq 0$) has a dramatic influence on the phase diagram obtained from the behaviour of the mean-field master equations in the long-time limit. In comparison to the $\gamma=0$ case, the phase diagram is enriched by the presence of oscillatory and bistable phases, see Fig.\,\ref{FIG:VGPhasesandOscillations} and Fig.\,\ref{Fig:DRPhaseDiagramkappagamma}.
We first turn our attention to the case where there is a small amount of dissipation in the atomic channel ($\gamma=0.01\Delta$) to analyse its impact on the long-time limit behaviour of the steady-state phases depicted in Fig.\,\ref{Fig:DRPhaseDiagramkappaonly}. We observe (see Fig.\,\ref{FIG:VGPhasesandOscillations}) that allowing for a small amount of dissipation, there are no stable steady-states that involve a broken lattice symmetry $T_{\rm lat}$. The only steady-states in the investigated ($V/\Delta$,$g/\Delta$)-plane is the empty atom-cavity system $({\rm FP}_{\downarrow})$ and a uniform superradiant phase $({\rm SR_{UNI}})$. The remaining long-time limit behavior is characterized by persistent oscillations that can be uniform $({\rm SR_{UNI}-OSC})$ or non-uniform $({\rm (AFM+SR_{UNI})-OSC})$. As the Rydberg interaction is conditioned on population in the upper state, the phase boundary of the empty atom-cavity system is independent of $V$. Formally, we obtain its phase boundary by inspecting the eigenvalues of the stability matrix corresponding to the fixed point $\left(\braket{a}=\braket{\sigma^x}=\braket{\sigma^y}=0,\braket{\sigma^z}=-1\right)$. The real part of at least one of the associated stability eigenvalues becomes positive when 
\begin{align}
\gamma  \kappa  \left((\gamma +2 \kappa )^2+4 \Delta ^2\right)^2-32 \Delta  \lambda ^2 \omega_0 (\gamma +2 \kappa )^2+8 \gamma  \kappa  \omega_0^2 (\gamma -2 \Delta +2 \kappa ) (\gamma +2 (\Delta +\kappa ))+16 \gamma  \kappa  \omega_0^4=0.
\label{EQ:FPStabilityLines}
\end{align}
Solving for g yields,
\begin{align}
g_{(crit,FP_{\downarrow})}=\frac{\sqrt{\gamma  \kappa } \sqrt{\left((\gamma +2 \kappa )^2+4 \Delta ^2\right)^2+8 \omega_0^2 (\gamma -2 \Delta +2 \kappa ) (\gamma +2 (\Delta +\kappa ))+16 \omega_0^4}}{4 \sqrt{2} \sqrt{\Delta  \omega_0 (\gamma +2 \kappa )^2}}.
\end{align}
Here, we observe that at $g=g_{(crit,FP_{\downarrow})}$ the associated linearized stability matrix of Eqs.~(\ref{EQ:Sigmax1}-\ref{EQ:PhotonTerm}) obtains a pair of purely imaginary complex conjugated eigenvalues which signalizes that the system changes into a limit cycle via a Hopf bifurcation. 
Limit cycles in driven-dissipative models have been observed before e.g.\, in spin-1/2 systems \cite{Chan15,Wilson16} and with Bose-Einstein condensates in optical cavities, see e.g. References \cite{Keeling10,Piazza15} and in driven QED-cavity arrays \cite{Jin14}.

The transition into the stable SR$_{\rm UNI}$-phase is discontinuous and we can solve  Eqs.~(\ref{EQ:Sigmax1}-\ref{EQ:PhotonTerm}) in the long-time limit to obtain the photon condensate as
\begin{align}
&|\braket{a}|^2\nonumber\\
&=\frac{16 J^3 (2 V-\Delta )\pm2 \sqrt{J^2 (\Delta +4 J)^2 \left(4 J^2 (\Delta -2 V)^2-2 \gamma ^2 J V-\gamma ^2 V^2\right)}-2 J^2 \left(\gamma ^2+2 \Delta  (\Delta -2 V)\right)-\gamma ^2 J V}{16 J \omega_0 (2 J+V)^2}\nonumber \\
&=\frac{J^2 (\Delta +4 J) (2 V-\Delta )\pm\sqrt{J^4 (\Delta +4 J)^2 (\Delta -2 V)^2}}{4 J \omega_0 (2 J+V)^2}+\mathcal{O}(\gamma^2)\;,
\label{EQ:VGUniFormPhaseEq}
\end{align}
where the coupling constant $J(\kappa,\omega_0)$ is given by Eq.\,\eqref{EQ:PhotonicJ}. Here the $\pm$ sign indicates that there are two branches of which only one is stable. Stable solutions (SR$_{\rm UNI}$) are depicted in Fig.\,\ref{subfig:PhaseDiagramVGsmallgamma}. In the limit of weak atomic noise, it can be seen from Eq.\,(\ref{EQ:VGUniFormPhaseEq}) that a stable solution to zeroth-order in $\gamma$ must satisfy $V>\Delta/2$.
The phase boundary between the two oscillating phases depicted in Fig.\,\ref{FIG:VGPhasesandOscillations} 
is obtained by comparing the oscillation amplitudes (in Fig.\,\ref{subfig:OscillationsVGPhases}) 
on the even vs.\, the odd sublattices in the long-time limit that result from direct integration of Eqs.~(\ref{EQ:Sigmax1}-\ref{EQ:PhotonTerm}). 

\begin{figure}
{\includegraphics[width=100mm]{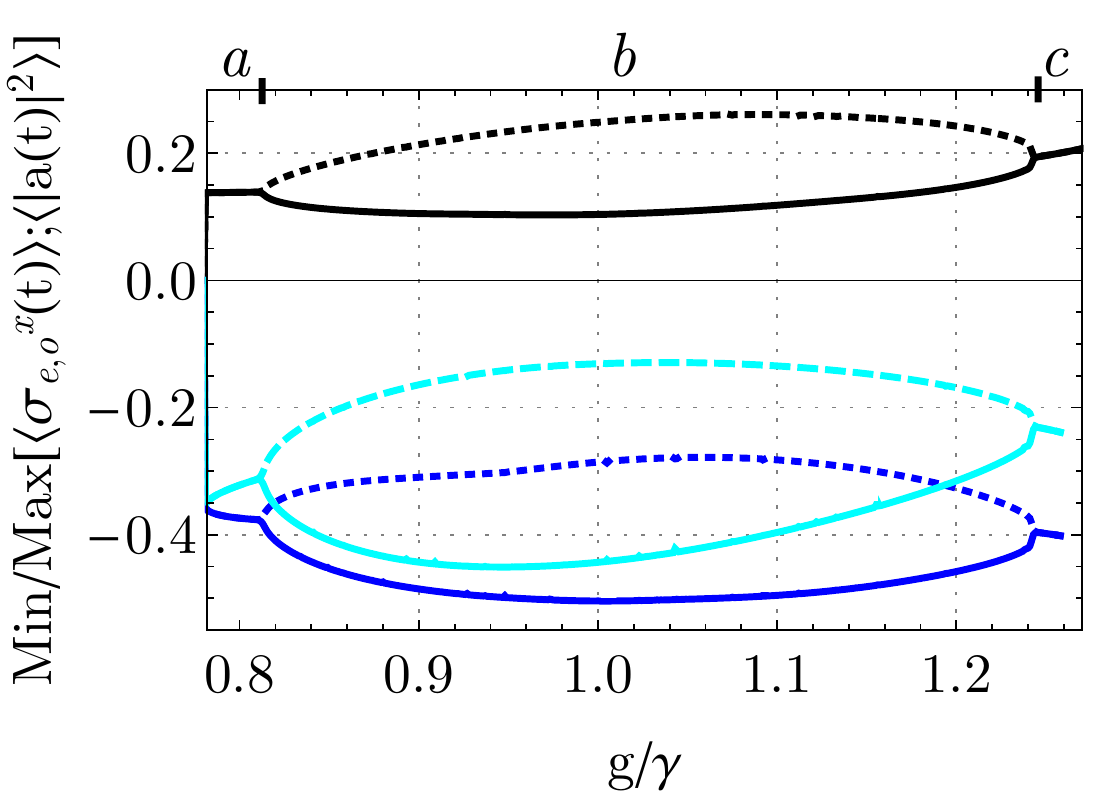}}
%\subfloat[]{\includegraphics[width=8cm]{AmpOscillations.pdf}\label{subfig:OrderParameterskappagamma}}\\
\caption{Amplitudes of the oscillations in the (AFM+SR)-OSC phase. Dashed (solid) blue lines show max(min)$[\langle\sigma^x_{e,o}(t)\rangle]$ and black lines show max(min)$[\langle |a(t)|^2 \rangle]$. Data is obtained by extracting the minimum and maximum of the amplitudes of $\langle\sigma^x_{e,o}(t)\rangle$ and $\langle |a(t)|^2 \rangle$ in a time interval chosen such that it contains several oscillations (if any are present) at long times. If the minimum and maximum coincide, the system settled into a steady state ($a,c$) corresponding to the (AFM+SR) phase, otherwise the system is in the (AFM+SR)-OSC limit cycle phase ($b$). Close to the (AFM+SR)-phase, the amplitudes decay continuously.  Parameters: $(\omega_0/\gamma=2.0, \Delta/\gamma=0.15, \kappa/\gamma=0.2, V/\gamma=1.8)$}
\label{SubFig:AmpOscillations}
\end{figure}
Next, we turn our attention to the features of the phase diagram depicted in Fig.\,\ref{Fig:DRPhaseDiagramkappagamma} where losses in the atomic channel can be strong. 
We analyze the oscillations of both atomic and photonic components in the long-time limit by explicit integration of Eqs.~(\ref{EQ:Sigmax1}-\ref{EQ:PhotonTerm}). Numerically we find persistent oscillations close to the (AFM+SR) region that are different on the even/odd sublattice, see Fig.\,\ref{fig:oscillations}. We determine the phase boundary for the(AFM+SR)-OSC phase depicted in Fig.\,\ref{Fig:DRPhaseDiagramkappagamma} by sampling initial conditions for the atomic components on the Bloch sphere and then integrating the set of Eqs.~(\ref{EQ:Sigmax1}-\ref{EQ:PhotonTerm}) directly. The phase boundary is set by the parameters ($\Delta/\gamma,g/\gamma$) for which the long-time limit is determined by the empty atom-cavity system (FP$_{\downarrow}$) for all initial conditions. 

\begin{figure}
%\subfloat[]{\includegraphics[width=8cm]{DRPhaseDiagram.pdf}\label{subfig:DRPhaseDiagramkappagamma}}
\subfloat[]{\includegraphics[width=90mm]{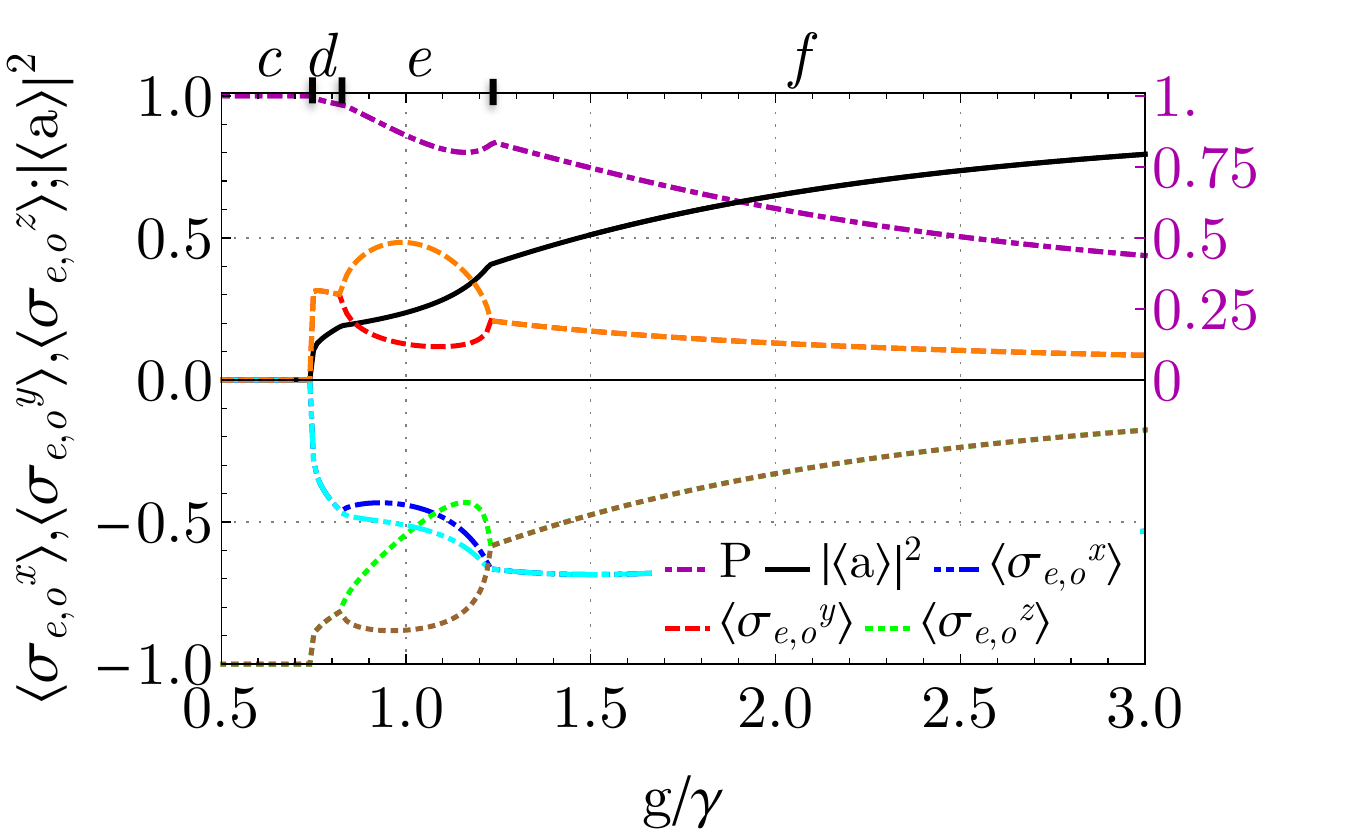}\label{subfig:OrderParameterskappagamma}}
\subfloat[]{\includegraphics[width=80mm]{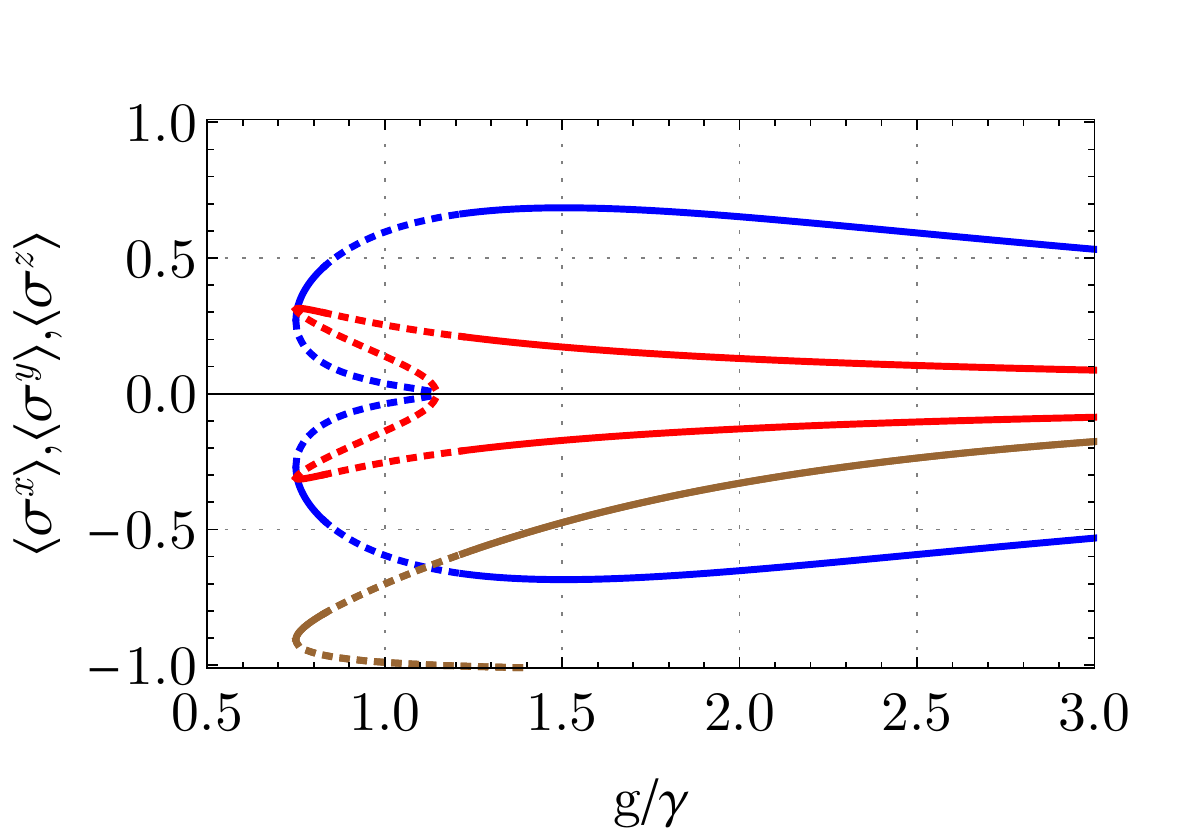}\label{subfig:HomogeneousStabilityAnalysis}}\\
\caption{Non-monotonous behaviour of the order-parameters as the atom-light coupling $g/\gamma$ is varied. The system changes discontinuously from the empty atom-cavity system (FP$_{\downarrow}$,$c$) into a homogeneous phase (${\rm SR_{UNI}}$,$d$) that becomes unstable towards an (AFM+SR,$e$) phase that disappears again in favor for an (${\rm SR_{UNI}}$,$f$) phase. On the right axis, the purity is shown that consistently decays, indicating the transition into a mixed state. (b) Stability analysis for homogeneous solutions as plotted in (a). (Unstable) stable, homogeneous solutions are plotted as (dotted) thick lines. The transition from the FP$_{\downarrow}$ state into the ${\rm SR_{UNI}}$ state is discontinuous if $V/\gamma>0$.  Parameters: $(\omega_0/\gamma=2.0, \Delta/\gamma=-0.1, \kappa/\gamma=0.2, V/\gamma=1.8)$}
\label{FIG:MagneticOrderingwithgamma}
\end{figure}

In Fig.~\ref{SubFig:AmpOscillations} we track the behaviour of the amplitude of the oscillations as a function of $g/\gamma$ and observe that the amplitudes decay continuously as the (AMR+SR) phase is approached. Numerically, we find no evidence that the (AFM+SR)-phase becomes unstable towards Hopf bifurcations meaning that stable limit cycles occur only outside the AFM+SR phase.

We continue our analysis by describing the behaviour of the magnetisations in the different domains of the phase diagram depicted in Fig.\,\ref{Fig:DRPhaseDiagramkappagamma}.
In Fig.\,\ref{subfig:OrderParameterskappagamma} we plot the magnetisation values in the steady-state for increasing the atom-light coupling $g/\gamma$. Starting in the empty atom-cavity, the system changes discontinuously into a ${\rm SR_{\rm UNI}}$ phase that soon after becomes unstable towards even/odd sublattice magnetisations (AFM+SR) that disappear again in favour for a re-entrance of the ${\rm SR_{\rm UNI}}$ phase. We depict the homogeneous solutions ${\rm SR_{\rm UNI}}$ and their stability in Fig.\,\ref{subfig:HomogeneousStabilityAnalysis}. We find that for $V/\gamma > 0$ the transition into the Dicke superradiance state is discontinuous.

We note that with $\gamma \neq 0$ the length of the semi-classical Bloch vector $\braket{\b{S}_{e,o}}=\l\braket{\sigma^x_{e,o}},\braket{\sigma^y_{e,o}},\braket{\sigma^z_{e,o}}\r$ is not conserved any more and can shrink for increasing $g/\gamma$ values. In equilibrium systems, an increase in the coupling parameter should stabilize the order in the steady-state, here we instead observe a non-monotonous behaviour where the 'order parameters' decay again after having reached a maximum value. We illustrate this decay by plotting the purity $P={\rm Tr}[(\rho_{e}\otimes \rho_o)^2]={\rm Tr}[\rho_{e}^2]{\rm Tr}[\rho_{o}^2]$ of the density matrix alongside the magnetisations. Both quantities decay as $g/\gamma$ is increased. In the case of the purity $P$ this indicates the decay towards a purely mixed state. 

The phase transitions in and out of the (AFM+SR) phases are continuous, whereas transitions from the empty atom-cavity system into the ${\rm SR_{UNI}}$ phase are discontinuous for $V/\gamma>0$, see Fig.\,\ref{FIG:MagneticOrderingwithgamma}.
On a mean-field level we observe bistabilities in the phase diagram depicted in Fig.\,\ref{Fig:DRPhaseDiagramkappagamma}. These could be induced by the nonlinearities in the mean-field master equations or can hint at non-trivial behaviour induced by fluctuations where the system in the long-time limit switches between the two steady-states predicted on a mean-field level \cite{tony12}. Mostly, bistabilities occur with the empty atom-cavity system (FP$_{\downarrow}$). The corresponding stability line can be calculated analytically from the stability matrix and we find that it is independent of $V$ since the Rydberg-dressed interaction is conditioned on population in the excited state, see Eq.\,\eqref{EQ:FPStabilityLines}. The size of the (AFM+SR) region instead does depend on $V$.

\subsection{Discussion of beyond mean-field effects}
\label{subsubsec:beyond}

Our analysis is based on mean-field theory and in this section we briefly discuss effects not captured by our approach and alternative theoretical approaches used in the literature. 
One possibility to capture fluctuations in 
far-from-equilibrium quantum spin systems is the real-time quantum field theory approach
\cite{schad15,babadi15} representing the spins as (Majorana) fermions. The number of fermion ``flavors'' 
and possibly constraints/gauge fields make this a complex endeavour, but, once developed, 
will be capable of treating even very large systems.

% Typically, these models focus on short-range couplings. The model that was considered here features interacting qubits that % couple to a single mode of a dynamical light-field and thus includes both short and long-range couplings. Consequently, the % model we consider is affected by the build up of short and or long-range correlations, quantum fluctuations between the 
% spins and the dynamics of fluctuating light fields. 
% A qualitative assessment of these beyond mean-field effects is non-trivial and beyond the scope of the present work.
%This will also help to place our study in its appropriate research context. 
% In equilibrium systems, mean-field calculations are an indispensable tool for predicting phase transitions and phase %diagrams. 

Several studies investigated effects beyond mean field in driven-dissipative lattice models 
that allow to acquire some intuition for the effect of correlations and fluctuations and for the validity of single-site mean-field studies in driven-lattice models out of equilibrium. Specifically, various numerical techniques such as 
variational approaches \cite{Weimer151,Weimer152}, cluster models \cite{Jin14,Jin16}, 
matrix product states \cite{Angelakis16}, or quantum trajectories \cite{tony12,tony132}
can help to shed light on the effects of correlations and fluctuations in the steady-state. 
However, most of these studies have focussed on a single short-range interaction.
%Recall that numerical approaches are complementary to the analytical field theory approaches in the 
%sense that they can often treat higher orders for limited system size, where the latter have the advantage 
%of expanding around infinite systems.
%Whereas for a driven-dissipative spin 1/2 XYZ-chain a cluster mean-field approach could reveal changes in 
%topology  in the phase diagram \cite{Jin16} (which is rarely, if ever, seen in equilibrium), a bosonic system of 
%a driven QED cavity array with cross-Kerr nonlinearities was well described in terms of a single
%-site mean field study with cluster techniques leading to minor modifications \cite{Jin14}.
Using these techniques, it was, for example, found that bistable regions may become washed out, when some form of correlation is taken into account \cite{Maghrebi16,Weimer151,Weimer152}. 

%In low-dimensional systems it is possible to compare mean-field studies with numerical calculations. Here, matrix product %states or quantum trajectories \cite{tony12,tony132} methods are applied to capture correlations of the system. In both of the %aforementioned papers, regions of bistability on a mean-field level hinted at non-trivial dynamics in the long-time limit and %thus served as an important guiding tool to uncover new effects. In the second paper, the quantum model exhibited collective %jumps between the two classical stable states and in the first paper, mean-field bistable regions where broken by correlations %and a new bunching antibunching regime emerged. The comparison of mean-field calculations with numerical methods is %strongly limited to low-dimensional and often to one-dimensional systems, especially when long-range correlations are to be %taken into account which is why quantum simulators are proposed to model driven-lattice models \cite{Angelakis16}.

%As we mention above, cluster mean-field, variational approaches and matrix product states in tensor network theory might %allow for a first insight into the effects of short-range correlations and quantum fluctuations in one-dimension and for very %small system sizes in two dimensions. 

The most dramatic modifications of the mean-field behaviour can be expected in one-dimensional systems.
First, the loss processes will lead to a 
noise-induced effective temperature for the atomic spins \cite{dalla13,gelhausen16} and 
modify the distribution function of the atoms; to capture the latter effect a quantum kinetic 
analysis is necessary \cite{piazza14b}. 
In the presence of an effective temperature, it is known since Polyakov \cite{polyakov87} 
that instantons prohibit one-dimensional Ising systems from ordering on longest scales. 
This implies, that we expect that the AFM phases in Fig.~\ref{Fig:DRPhaseDiagramkappaonly}
will disappear for a 1D spin array (whether they will be replaced by SR phases or the vacuum depends 
on the sign of the detunings). As in 1d quantum systems, also here despite the absence of true 
long-range order, even-odd correlations will be visible in the correlation functions and structure factors, whose 
computation, however, can get quite involved/is not known in the absence of equilibrium. 
The order parameter of the SR phases will of course remain stable, due to
the effectively infinite range of the interaction, dimensionality is irrelevant here.

Our analysis was, however, focusing on a two-dimensional setup where not only SR but also the AFM order can be stable in the presence of thermal and non-equilibrium fluctuations as domain walls cost an infinite amount of energy.
For the two-dimensional spin array discussed in the present paper, we believe 
the qualitative features are robust, i.e. fluctuations might shift the phase transitions, induce 
a finite effective temperature, and 
further broaden the spectra but will not fully destroy the order. In particular the AFM+SR strip 
ending in the 
multi-critical point shown in Fig.~\ref{Fig:DRPhaseDiagramkappaonly} might get washed out and/or replaced by a first order transition. Mean-field also does not capture all critical properties. Note that not much is known about first-order transitions in multi-critical, 
driven-dissipative systems. Perhaps the most difficult question is whether the oscillating phases shown in the phasediagram of Fig.~\ref{Fig:DRPhaseDiagramkappagamma} will survive fluctuations beyond mean field. Here it is important to note that due to the infinite range interaction mediated by the cavity, a spontaneous breaking of time-translational symmetry is possible in two (and even one) dimension and is not destroyed by 
goldstone-mode fluctuations \cite{tony13}. Nevertheless, it is unclear whether the heating processes naturally occuring in periodically driven interacting many-particle systems \cite{maxGenske} will be compensated by cooling processes due to radiative losses in such a way that the oscillations survive.

%\newpage
\section{Conclusions and future directions}
\label{sec:conclusions}
The point of this paper was to create a base case (model and approximate solution) for a 
large array of \emph{self-interacting atomic qubits} coupled to a single-mode optical light field. 
Why do we believe this is needed? Because there is an increasing number of 
experimental platforms ranging from ultracold atoms in optical cavities, superconducting circuits, 
photonic pulses travelling through Rydberg gases to nano-photonic crystals, seeking to scale up the 
number of qubits and interface them with photons. 
Qubit-qubit interactions can be wanted --to mediate photon-photon 
nonlinearities for example-- or stray, in which case they would be seen to dephase 
a collective coupling of a set of qubits to photons. In the face of the considerable complexity these systems 
generate --large number of quantum spins, fluctuating light fields, non-equilibrium aspects--
our simple model has yielded some experimentally directly testable predictions: A regime where magnetic translation 
symmetry breaking and superradiance occur together, a new even-odd collective mode in the cavity spectrum, 
and intriguing, oscillating solutions for both, the spin components and a coherent photon field.
Present-day technology with
Rydberg-dressed spin lattices in optical cavities should be able to check and refine these results.
%\subsection{Future directions}
%
Unfortunately, we were not able to solve even our simple model exactly; the 
Rydberg-mediated nearest-neighbour interaction does induce non-trivial quantum fluctuations (centered 
around the even-odd modulation momentum $(\pi,\pi)$) between the spins.
Our non-equilibrium mean-field ansatz for the density matrix kept track of only 
the expectation values of the spin components on the even and odd sub-lattices, and the photon field, 
respectively. We are not aware of a developed technique (see appendix \ref{subsubsec:beyond} for a brief discussion), which can capture
quantum fluctuations for large, far-from-equilibrium quantum spin systems coupled 
to the (potentially large) Hilbert space of one or multiple photon modes. 
Promising efforts in this direction invoke a fermion 
representation of the quantum spins on the closed-time Keldysh contour 
(\cite{schad15,babadi15} and references therein); this then, in principle, allows to leverage 
over diagrammatic techniques well-developed for ground state fermions. Work in this direction 
is underway.
Particularly promising physical set-ups to study the interplay of interacting 
qubits with light in the future are nano-photonic and 
one-dimensional quantum-optical systems 
\cite{thompson12,chang13,rauschenbeutel14, ramos14,pichler15,asenjo16, douglas15,thompson15}, in which 
huge effects from even small qubit-qubit interactions can be expected. Moreover, these 
systems typically contain an (infinite) continuum of photon modes significantly enriching 
the complexity of the photonic Hilbert space at one's disposal.
The same is true for optical resonators in multi-mode operation 
\cite{ningyuan16,schmitt15, kollar16} making them also interesting 
targets for further explorations.

\acknowledgments
We thank A. Gl\"atzle, C. Kollath, and  P. Zoller for good discussions and 
further acknowledge helpful remarks by T. Pohl, P. Rabl, H. Ritsch, and 
J. Zeiher during the Quantum Optics 2016 conference. 
JG thanks the Department of Physics at Harvard University for hospitality, where 
this research was partly carried out (during the first half of 2015). We 
are grateful to M. D. Lukin for discussions on related topics. 
%We also thank the Sachdev Group for sharing their bibliography style file.
This work was supported by the Leibniz Prize of A. Rosch, by the 
Harvard-MIT Center for Ultracold Atoms (CUA), by the Multidisciplinary University Research Initiative (MURI), 
and by the German Research Foundation (DFG) through CRC 1238 and
the Institutional Strategy of the University of Cologne within the 
German Excellence Initiative (ZUK 81).

%%%%%%%%%%%%%%%%%%%%%%%%%%%%%%%%%%%%%%%%%
\appendix

\section{Mean-field solution of the T=0 equilibrium spin model}
\label{Sec:Equilibrium}

In this section, we analyze Hamiltonians $H_{\rm spin-light}$ Eq.\,\eqref{eq:H_spin-light} and $H_{\rm spin}$ 
Eq.\,\eqref{eq:H_spin} within a (standard) equilibrium mean-field theory for spins.
We will find similar phases to those in Fig.\,\ref{Fig:DRPhaseDiagramkappaonly}  upon 
identifying one of the spin-spin interaction constants with cavity parameters, somewhat surprisingly 
including the photon decay $\kappa$. The deeper reason for this is that with the counter-rotating terms, 
in the atom-light coupling the excitation number is stabilized despite the loss rates. Accordingly, the 
non-equilibrium steady-state phase in the long-time limit are then qualitatively similar to the ground state phases. 
Dynamics and statistics (effective temperature) of the full non-equilibrium system, however, remain
qualitatively drastically different.

First, we integrate out the quadratic photon terms which yields an effective Hamiltonian that features a ferromagnetic all-to-all atom-atom coupling $J$. The connection to the non-equilibrium system is then made explicit by letting $J$ depend on $\kappa$ as pointed out below.  The Hamiltonian we consider is written as
\begin{align}
\tilde{H}=-\frac{J}{N}\sum\limits_{i j}\sigma^{x}_{i}\sigma^{x}_{j}-\frac{\Delta}{2}\sum\limits_{i}\sigma^z_i+\frac{1}{2}V\sum_{\braket{\ell m}}^{N}\sigma^{ee}_{\ell}\sigma^{ee}_m
\end{align}
Where the $1/2$ in front of the Rydberg interaction avoids overcounting. We cast the last terms into a spin-language with the replacement $\sigma^{ee}_{\ell}=1/2 (1+\sigma^z_{\ell})$.
We decouple the interaction terms in mean-field theory by expanding the operators around their mean-value to linear order in fluctuations. We neglect all second-order fluctuation terms and write the effective spin-Hamiltonian in a form that resembles the interaction of the spin-variables with an effective, local magnetic field that needs to be determined self-consistently and represents the mean field from the neighbouring spins.
Ignoring constant energy shifts, the full mean-field Hamiltonian assuming $d=2$-spatial dimensions is given as
\begin{align}
\tilde{H}^{\rm MF}&=-V \frac{N}{2} \braket{\sigma_{\rm even}^z}\braket{\sigma_{\rm odd}^z}+NJ \braket{\sigma^x}^2+\sum\limits_{i \in even}\b{B}^{\rm even}\b{\sigma}_{i}
+\sum\limits_{j \in odd}\b{B}^{\rm odd} \b{\sigma}_{j}
\end{align}
Here, we have already accounted for an $even/odd$ sub-lattice asymmetry in the $z-$components.
We use the vector of Pauli matrices as $\b{\sigma}=\left(\sigma^x,\sigma^y,\sigma^z\right)^T$ and define the local magnetic fields as
\begin{align}
\b{B}^{\rm even/odd}=\left(\bigg[ \braket{\sigma_{\rm odd/even}^z}V+\left(V-\Delta /2\right)\bigg]\b{\hat{z}}+2J\braket{\sigma^x}\b{\hat{x}}\right)
\end{align}
We evaluate the partition sum
\begin{align}
Z&=\text{Tr}\bigg[e^{-\beta \tilde{H}^{\rm MF}}\bigg]=2^N\bigg[\cosh\left(|\b{B}^{\rm even}|\right)\cosh\left(|\b{B}^{\rm odd}|\right)\bigg]^{N/2}\exp\bigg[{V\frac{N}{2}\beta \braket{\sigma_{\rm even}^z}\braket{\sigma_{\rm odd}^z}-N \beta J \braket{\sigma^x}^2}\bigg]
\label{EQ:PartitionFunction}
\end{align}
to obtain the self-consistency equations for the order-parameters
\begin{align}
\phi&=\frac{\braket{\sigma^x_{\rm even}}+\braket{\sigma^x_{\rm odd}}}{2}=\frac{1}{2}\tanh\left(\beta|\b{B}^{\rm odd}|\right)\frac{B^{\rm odd}_{x}}{|\b{B}^{\rm odd}|}+\frac{1}{2}\tanh\left(\beta|\b{B}^{\rm even}|\right)\frac{B^{\rm even}_{x}}{|\b{B}^{\rm even}|}\label{EQ:EquilibriumOrder1}\\
\rho &=\frac{\braket{\sigma_{\rm even}^z}-\braket{\sigma_{\rm odd}^z}}{2}=\frac{1}{2}\tanh\left(\beta|\b{B}^{\rm even}|\right)\frac{B^{\rm even}_{z}}{|\b{B}^{\rm even}|}-\frac{1}{2}\tanh\left(\beta|\b{B}^{\rm odd}|\right)\frac{B^{\rm odd}_{z}}{|\b{B}^{\rm odd}|}\\
\rho_0 &=\frac{\braket{\sigma_{\rm even}^z}+\braket{\sigma_{\rm odd}^z}}{2}=\frac{1}{2}\tanh\left(\beta|\b{B}^{\rm even}|\right)\frac{B^{\rm even}_{z}}{|\b{B}^{\rm even}|}+\frac{1}{2}\tanh\left(\beta|\b{B}^{\rm odd}|\right)\frac{B^{\rm odd}_{z}}{|\b{B}^{\rm odd}|}\label{EQ:EquilibriumOrder3}
\end{align}
Where $\rho$ is the staggered magnetisation and $\rho_0$ is the average magnetisation in the $z$ direction. The magnetic order parameter $\phi$ measures the magnetisation in $x$-direction and $\beta=1/T$ is the inverse temperature.
We denote the free energy per spin in the zero temperature limit $T \to 0$ as
\begin{align}
f=\left. \frac{F}{N}\right|_{T\to 0}&=\left. -\frac{T}{N}\log(Z)\right|_{T\to 0}=\frac{1}{2} V \left(\rho ^2-\rho_0^2\right)+J \phi ^2\nonumber \\
&-\frac{1}{2} \left(\sqrt{\left(V (\rho_0-\rho )+V-\frac{\Delta }{2}\right)^2+4 J^2 \phi ^2}+\sqrt{\left(V (\rho +\rho_0)+V-\frac{\Delta }{2}\right)^2+4 J^2 \phi ^2}\right)
\label{EQ:FreeEnergyZeroT}
\end{align}
We can determine the zero-temperature phase-diagram by solving the coupled set of Eqs.$~\left(\ref{EQ:EquilibriumOrder1}-\ref{EQ:EquilibriumOrder3}\right)$ numerically and retain only the solutions with the lowest free-energy according to \eq{EQ:FreeEnergyZeroT}. We obtain the splitting in the $(\braket{\sigma^x_{\rm even}},\braket{\sigma^x_{\rm odd}})$ components by using Eq.\,\eqref{EQ:PseudoangularmomentumConstraint}. We find that we can map the equilibrium phase-diagram to the phase diagram obtained by calculating the non-equilibrium steady-states (see Fig.\,\ref{Fig:DRPhaseDiagramkappaonly}) if we identify the ferromagnetic exchange coupling as
\begin{align}
J(g,\kappa)=\frac{g^2\omega_0}{\omega_0^2+\kappa^2}
\label{EQ:PhotonicJ}
\end{align}
This coupling is inferred from solving for the steady-state values of the photons (see \eq{EQ:PhotonTerm}) which is given as $g\left(\braket{a}+\braket{a^{\dagger}}\right)=-\frac{1}{2}g\braket{\sigma^x_{\rm even}+\sigma^x_{\rm odd}}\left(\frac{g}{\omega -i \kappa }+\frac{g}{\omega +i \kappa }\right)\propto J(g,\kappa)$. Allowing the interpretation that the photonic losses with rate $\kappa$ weaken the atom-atom couplings.

%%%%%%%%%%%%%%%%%%%%%%%%%%%%%%%%%%%%%%%%%%%%%%%%%%%%%%%%%%%%%%%
\section{Transformation of fully time-dependent model into rotating frame}
\label{sec:DHDR}

Here, we detail calculations where we derive how the parameters of the Hamiltonians $H_{\rm spin-light}$ given by Eq.\,\eqref{eq:H_spin-light} and $H_{\rm spin}$ given by Eq.\,\eqref{eq:H_spin} are related to tunable laser parameters that result from the optical implementation shown in Fig.\,\ref{Fig:RydbergDressingScheme}. The Hamiltonian we consider is of the form
\begin{align}
H&=H_{\rm cav}+H_{\rm atoms} +H(t)_{\rm pump}+H_{\rm atom-light}+H_{\rm atom-atom}\\
H_{\rm cav}&=\omega_0 a^{\dagger}a\\
H_{\rm atoms}&=\sum\limits_{\ell=1}^N \omega_d \ket{d}_{\ell}\bra{d}+\omega_e \ket{e}_{\ell}\bra{e}+\omega_{\rm Ryd}\ket{\rm Ryd}_{\ell}\bra{\rm Ryd}+\omega_1 \ket{1}_{\ell }\bra{1}\\
H_{\rm pump}(t)&=\sum\limits_{\ell=1}^N\frac{\Omega_e}{2}e^{-i \omega_{\Delta e}t}\ket{e}_{\ell}\bra{1}+\frac{\Omega_d}{2}e^{-i \omega_{\Delta d}t}\ket{d}_{\ell}\bra{0}+\frac{\Omega_{\rm Ryd}}{2}e^{-i \omega_{\Delta r}t}\ket{\rm Ryd}_{\ell}\bra{1}+{\rm h.c.}\\
H_{\rm atom-light}&=\sum\limits_{\ell=1}^N\l g_{d}\ket{d}_{\ell}\bra{1}+g_{e}\ket{e}_{\ell}\bra{0}\r a+{\rm h.c.}\\
H_{\rm atom-atom}&=\sum\limits_{\braket{\ell m}}V_{\ell m}\l \ket{\rm Ryd}_{\ell}\bra{\rm Ryd} \r \l  \ket{\rm Ryd}_{m}\bra{\rm Ryd} \r
\end{align}
The frequencies $(\omega_d, \omega_e, \omega_{\rm Ryd},\omega_1)$ refer to the atomic levels labelled by the sequence $(d,e,{\rm Ryd},1)$ and are measured relative to the atomic level $\ket{0}$. Correspondingly, the frequencies $(\omega_{\Delta d}, \omega_{\Delta e}, \omega_{\rm \Delta r})$ refer to the laser frequencies of the pump-terms.  Here, $\omega_0$ denotes the bare cavity resonance. We have assumed homogeneous pumping of the atoms from the side $\Omega_{(d,e);\ell}\approx \Omega_{(d,e)}$ and a homogeneous coupling of the light field to the atoms $g_{(d,e);\ell}\approx g_{(d,e)}$. We eliminate the explicit time-dependence by switching into a rotating frame such that the new Hamiltonian reads
\begin{align}
\tilde{H}&=U^{\dagger}H_0U-U^{\dagger}i\partial t U
\end{align}
where the Hamiltonian $H_0$ is given as
\begin{align}
U(t)&=\exp\l -i H_0 t\r\\
H_0&= \l \omega_{\Delta d}-\omega_1' \r  a^{\dagger}a+\sum\limits_{\ell=1}^N \bigg[\l \omega_{\Delta e}+\omega_1' \r \ket{e}_{\ell}\bra{e}+\omega_{\Delta d}\ket{d}_{\ell}\bra{d} +\omega_1'\ket{1}_{\ell}\bra{1}+\left( \omega_{\rm Ryd}+\omega_1'\right) \ket{\rm Ryd}_{\ell}\bra{\rm Ryd}\bigg]
\end{align}
Cross coupling lasers need to be tuned such that they are strongly detuned from the levels $(\ket{d},\ket{e})$ which can then be eliminated adiabatically. Under the condition $|\Delta_{d,e}| \gg \kappa, \gamma, \Omega_{d,e}$, the dynamics of the system are now described by an effective Hamiltonian $\tilde{H}=\tilde{H}_{\rm Ryd}+\tilde{H}_{\rm 10}+\tilde{H}_{\rm L}$:
\begin{align}
\tilde{H}_{L}&=\sum_{\ell=1}^N \frac{\Omega_{\rm Ryd}}{2}\l  \ket{\rm Ryd}_{\ell}\bra{1}+\ket{1}_{\ell}\bra{\rm Ryd} \r\\
\tilde{H}_{\rm Ryd}=&-\Delta_{\rm Ryd}\sum\limits_{\ell=1}^N \ket{\rm Ryd}_{\ell}\bra{\rm Ryd}+ \sum\limits_{\braket{\ell m}}V_{\ell m}\l \ket{\rm Ryd}_{\ell}\bra{\rm Ryd} \r \l  \ket{\rm Ryd}_{m}\bra{\rm Ryd} \r\\
\tilde{H}_{\rm 10}=&\omega_a a^{\dagger}a+\sum_{\ell=1}^N\bigg[\l \Delta_1 +\frac{\Omega^2_e}{4\Delta_e} \r \ket{1}_{\ell}\bra{1}  + \frac{\Omega_d}{4\Delta_d}\ket{0}_{\ell}\bra{0}+\frac{1}{2}\l \frac{g_e \Omega_e}{\Delta_e} \ket{0}_{\ell}\bra{1}a^{\dagger} + \frac{g_d \Omega_d}{\Delta_d} \ket{1}_{\ell}\bra{0} a^{\dagger} + {\rm h.c.} \r \label{EQ:H0inrotframe}\\
&+\l \frac{g^2_e}{\Delta_e} \ket{0}_{\ell}\bra{0} + \frac{g^2_d}{\Delta_d} \ket{1}_{\ell}\bra{1}\r  a^{\dagger}a\bigg] \nonumber
\end{align}
where we have used the following frequencies
\begin{align}
\omega_a&=\omega_0-\l \omega_{\Delta d}-\omega_1' \r, \nonumber\\
\quad\Delta_{\rm Ryd}&=-[\omega_{\rm Ryd}-\l \omega_{\Delta r}+\omega_1' \r] \nonumber\\
\Delta_1&=\omega_1-\omega_1',\quad 
\nonumber\\
2\omega_1'&=\omega_{\Delta d}-\omega_{\Delta e}\;.
\label{EQ:Rotatingfrequencies}
\end{align}
In a next step, high-lying Rydberg states are admixed to the ground-states $\ket{1}_{\ell}$ to realise a Rydberg-dressed interaction between the states $\ket{\tilde{1}}=\ket{1}+\frac{\Omega_{\rm Ryd}}{2\Delta_{\rm Ryd}}\ket{\rm Ryd}+\mathcal{O}\l\frac{\Omega_{\rm Ryd}}{2\Delta_{\rm Ryd}}\r$. Typically,  two-body Born-Oppenheimer potentials as a function of the distance $r_{ij}$ between two Rydberg levels are obtained by diagonalising Hamiltonians of the form $H_{L}+H_{\rm Ryd}$ in a two-atom basis \cite{GlaetleZoller12}. A detailed calculation that includes coupling to the complicated level structure is thus highly non-trivial.  Focusing on the weak-dressing regime $\Omega_{\rm Ryd}/ \Delta_{\rm Ryd}\ll 1$ and red-detuning of the dressing laser we follow the many-body perturbation expansion performed in Ref.\,\cite{Pohl10} to obtain
the effective Hamiltonian for the Rydberg part to leading order in the corrections
\begin{align}
\tilde{H}_{\rm Ryd}&=-\frac{\Omega^2_{\rm Ryd}}{4 \Delta_{\rm Ryd}}\sum\limits_{\ell=1}^N  \ket{\rm \tilde{1}}_{\ell}\bra{\rm \tilde{1}} +\frac{1}{2}\l\frac{\Omega_{\rm Ryd}}{2\Delta_{\rm Ryd}}\r^4\sum\limits_{i \neq j} \frac{C_6}{r_{ij}+R^6_c} \l \ket{\rm \tilde{1}}_{i}\bra{\rm \tilde{1}} \r \l  \ket{\rm \tilde{1}}_{j}\bra{\rm\tilde{1}}  \r \nonumber\\
&=-\frac{\Omega^2_{\rm Ryd}}{4 \Delta_{\rm Ryd}}\sum\limits_{\ell=1}^N  \ket{\rm \tilde{1}}_{\ell}\bra{\rm \tilde{1}} +\frac{1}{2}\sum\limits_{i \neq j} V^{\rm eff}_{\rm ij} \l \ket{\rm \tilde{1}}_{i}\bra{\rm \tilde{1}} \r \l  \ket{\rm \tilde{1}}_{j}\bra{\rm\tilde{1}}  \r 
\end{align}
It can be seen that the dressed states $\ket{\tilde{1}}$ acquire additional light-shifts $\sim \Omega_{\rm Ryd}^2/ (4 \Delta_{\rm Ryd})$ and the Rydberg potential is tunable by changing $(\Omega_{\rm Ryd},\Delta_{\rm Ryd})$. Here, $V^{\rm eff}_{ij}$ and $R_c$ are defined in Eq.\,\eqref{EQ:EffRydbergPotential}.
We now replace $\ket{1}$ with the dressed Rydberg state $\ket{\tilde{1}}$ everywhere in Eq.\,\eqref{EQ:H0inrotframe}. With
\begin{align}
\sum_{\ell=1}^N \ket{1}_{\ell}\bra{1}&=\frac{1}{2}\sum_{\ell=1}^N \l \ket{1}_{\ell} \bra{1}-\ket{0}_{\ell}\bra{0}+N \r=\frac{1}{2}\sum_{\ell=1}^N \sigma^z_{\ell}+\frac{N}{2}\\
\sum_{\ell=1}^N\ket{0}_{\ell}\bra{0}&=-\frac{1}{2} \sum_{\ell=1}^N\l \ket{1}_{\ell}\bra{1}-\ket{0}_{\ell}\bra{0}-N\r=-\frac{1}{2}\sum_{\ell=1}^N \sigma^z_{\ell}+\frac{N}{2}
\end{align}
the Hamiltonian is now cast into the form:
\begin{align}
\tilde{H}=&a^{\dagger}a\bigg[ \frac{N}{2} \l\frac{g^2_e}{\Delta_e}+\frac{g^2_d}{\Delta_d}\r +\omega_a \bigg]
\nonumber\\
&+
\sum_{\ell=1}^N\sigma^z_{\ell}\frac{1}{2} \bigg[\l \frac{\Omega^2_d}{4\Delta_d}-\frac{\Omega^2_e}{4\Delta_e}\r +\Delta_1-\frac{\Omega^2_{\rm Ryd}}{4\Delta_{\rm Ryd}} +\frac{1}{2}\sum_{m(m\neq \ell)}^NV^{\rm eff}_{m\ell}\bigg]
+\sum_{\ell=1}^N \sigma^z_{\ell}\frac{1}{2} \frac{1}{N}\l \frac{g^2_d}{\Delta_d}-\frac{g^2_e}{\Delta_e} \r a^{\dagger} a
\nonumber\\
&+ \sum_{\ell=1}^N \bigg[\frac{\lambda_d}{\sqrt{N}} \l \sigma^+_{\ell} a^{\dagger} + \sigma^-_{\ell} a  \r + \frac{\lambda_e}{\sqrt{N}} \l \sigma^+_{\ell} a + \sigma^-_{\ell} a^{\dagger}  \r\bigg]\nonumber\\
&
%\nonumber\\
%&
+ 
\frac{1}{2}\sum\limits_{i \neq j} V^{\rm eff}_{\rm ij} \frac{\sigma^z_i}{2} \frac{\sigma^z_j}{2}
\;,
\label{EQ:RotatingFrameH}
\end{align}
with effective spin-photon couplings (set equal and denoted by $g$ in Eq.~(\ref{eq:g}))
\begin{align}
\lambda_{d,e}&=\sqrt{N}\frac{g_{d,e}\Omega_{d,e}}{2\Delta_{d,e}}\;.
\end{align}
To generate AFM ordering it is advantageous for the effective longitudinal field corresponding to the second term in the first line in Eq.\,\eqref{EQ:RotatingFrameH} to be negative. We analyse typical orders of magnitudes.
 The hyperfine structure splitting in the ground state manifold is $\omega_1=2\pi \times 6.835$GHz.
Typically the cavity-assisted Raman transitions are achieved by coupling to the first excited state manifold that is split into a fine-structure $5^2P_{1/2}$ with $F'=2$ and $F'=1$ that for this choice is on the order of $812$MHz. The external driving lasers are separated by approximately twice the ground-state hyperfine splitting $\omega_{1'}=\frac{1}{2}\l \omega_{\Delta d}-\omega_{\Delta e} \r \sim \omega_1$ such that $\Delta_1=\omega_1-\omega_{1'}\sim$MHz. For weakly admixing the Rydberg state to the groundstate manifold $\ket{\uparrow}$ the detuning from the Rydberg state $\Delta_{\rm Ryd}$ must satisfy $\Omega_{\rm Ryd} \ll \Delta_{\rm Ryd}$. Typical Rabi frequencies for the drive to the Rydberg level are $\Omega_{\rm Ryd}\sim$MHz. The detuning from the Rydberg level now has a frequency component $\omega_{1'}$ from the Raman-scheme: $\Delta_{\rm Ryd}=-[\omega_r-(\omega_{\Delta r}+\omega_{1'})]$. This can take the usual detuning $\Delta_{\rm Ryd}$ far above the MHz regime which makes $\Delta_{\rm Ryd} \ggg \Omega_{\rm Ryd}$.
The longitudinal field $\left(\Delta_1-\frac{\Omega^2_{\rm Ryd}}{4\Delta_{\rm Ryd}}\right)$ for $\Omega_d=\Omega_e$ and $\Delta_d=\Delta_e$ is in the MHz range and can in principle be tuned positive and negative. 
%%%%%%%%%%%%%%%%%%%%%%%%%%%%%%%%%%%%%%%%%%%%%%%%%%%%%%%%%%%%%%%%%%%%%
\section{Validity analysis of the even-odd sublattice Ansatz}
\begin{figure}
{\includegraphics[width=100mm]{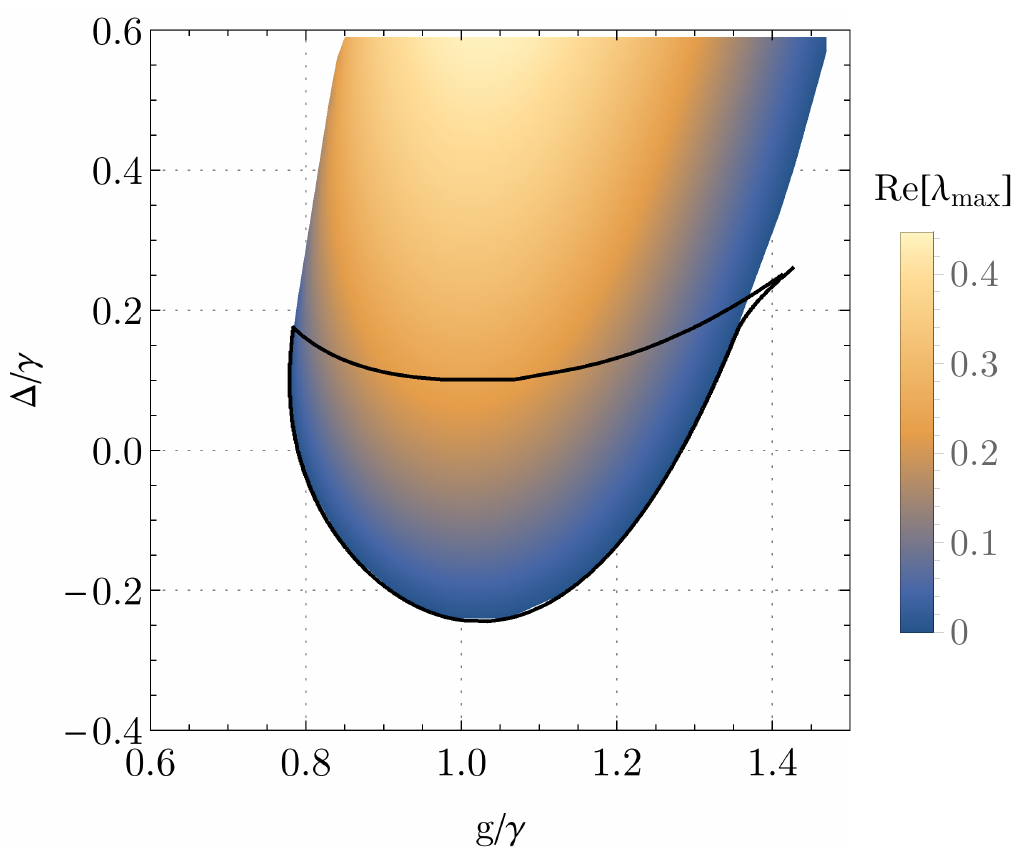}}
\caption{Instability of the homogeneous solution against excitations with wavevector $(k_x,k_y)=(\pi,\pi)$, calculated from Eq.\,\eqref{EQ:StabilityMatrixAppendix}. The color scale shows the real part of the eigenvalue that is maximally unstable. Using the even-odd sublattice Ansatz we find the phase-boundary of stable anti ferromagnetic solutions (enclosed by the bold line) which are consistent with the stability analysis of the homogenous solutions. In the upper half-plane, where $\Delta>0$, there is a region where the homogeneous solution is unstable against excitations with $(k_x,k_y)=(\pi,\pi)$ but the mean-field antiferromagnetic solutions are not stable above the bold line. The region around $(g/\gamma,\Delta/\gamma) \approx (1.4,0.18)$ is bistable and can show steady-states of $\rm{(AFM+SR)}$ or $\rm{SR_{UNI}}$ ordering. This plot is done for the same parameters as in the entire phase diagram that is given in Fig.\,\ref{Fig:DRPhaseDiagramkappagamma}}
\label{Fig:Stabilityanalysis}
\end{figure}

Here, we determine the linear stability of the homogeneous fixed points of Eqs.~(\ref{EQ:SigmaxApp}-\ref{EQ:adagApp}) against excitations with momentum $\b{k}$, see e.g. References \cite{Boite14,Wilson16}. In driven-dissipative lattice models with short-range interaction, orderings with incommensurate  wavevectors have been observed \cite{Wilson16,Boite13,Boite14,Schiro16,Zou14}. This can happen because of an interplay of dissipation and a competition of different, momentum-dependent interactions such as in a driven spin-1/2 XYZ-model \cite{tony13}. In driven-dissipative Bose-Hubbard-type lattice models (see e.g.\,Reference\,\cite{Zou14}) multimode photon fields are considered which have a finite momentum dependence. This is in contrast to our single-mode photon field that only couples to the zero-momentum component. The infinite-range atom-light and the antiferromagnetic spin-spin interaction suggest that the steady-states can either be uniform or that it can break the translational invariance of the system, respectively (see also Table \ref{Tab:OrderParameters}). We thus expect and find that homogeneous mean field solutions are maximally unstable either at $(k_x,k_y)=(0,0)$ or against excitations with $(k_x,k_y)=(\pi,\pi)$.   
We outline the analysis below.\\ On a mean-field level, we write down the master-equation for every lattice site $\b{n}$
\begin{align}
\partial_t\braket{\sigma^x_{\b{n}}(t)}=&\braket{\sigma^y(t)} [\Delta-\frac{V}{2}\sum_{\braket{\b{n}\b{m}}}( \braket{\sigma^z_{\b{m}}(t)} +1 )]-\frac{\gamma}{2}\braket{\sigma^x_{\b{n}}(t)} \label{EQ:SigmaxApp}\\
\partial_t\braket{\sigma^y_{\b{n}}(t)}=&\braket{\sigma^x_{\b{n}}(t)} [\frac{V}{2}\sum_{\braket{\b{n}\b{m}}}( \braket{\sigma^z_{\b{m}}(t)} +1 )-\Delta ]-2 g  [\braket{a(t)}+\braket{a^{\dagger}(t)}] \braket{\sigma^z_{\b{n}}(t)}-\frac{\gamma}{2}\braket{\sigma^y_{\b{n}}(t)} \\
\partial_t\braket{\sigma^z_{\b{n}}(t)}=&2 g  [\braket{a(t)}+\braket{a^{\dagger}(t)}]\braket{\sigma^y_{\b{n}}(t)}-\gamma(1+\braket{\sigma^z_{\b{n}}(t)})  \\
\partial_t\braket{a(t)}=&-(\kappa +i \omega_0)\braket{a(t)}- i g\sum_{\b{n}}\braket{\sigma^x_{\b{n}}(t)}\\
\partial_t\braket{a^{\dagger}(t)}=&-(\kappa -i \omega_0)\braket{a(t)}+ i g\sum_{\b{n}}\braket{\sigma^x_{\b{n}}(t)},
\label{EQ:adagApp}
\end{align}
where $\b{n}$ is a two-dimensional position vector on the square lattice.
We check the validity of our even-odd sublattice approach by adding plane-wave perturbations to the uniform steady-states with the Ansatz 
\begin{align}
\braket{\b{\sigma}_{\b{n}}(t)}&=\braket{\b{\sigma}}+\b{\delta \sigma}_{\b{n}}(t), \quad \braket{a(t)}=\braket{a}+\delta a(t)\;,
\end{align}
where $\b{\sigma}=(\braket{\sigma^x},\braket{\sigma^y},\braket{\sigma^z})^T$
are the homogeneous solutions to Eqs.~(\ref{EQ:SigmaxApp}-\ref{EQ:adagApp}) and $\b{k}$ contains the wave numbers of the perturbation.
We Fourier transform according to 
\begin{align}
\b{\delta \sigma_{n}(t)}=\frac{1}{N}\sum_{\b{k}}e^{i\b{k\cdot n}} \b{\delta \sigma}_{\b{k}}(t), \quad \b{k}=(k_x,k_y)^T, \quad k_{\rm \ell}=\frac{2\pi}{N}j, \quad j=0,\dots, N-1
\end{align}
We linearize equations Eqs.~(\ref{EQ:SigmaxApp}-\ref{EQ:adagApp}) in the fluctuations $(\b{\delta \sigma_{k}}(t),\delta a(t))$ and obtain a set of equations for each wave-vector $\b{k}$
\begin{align}
\partial_t \b{\delta}_{\b{k}}(t)&=\mathcal{D}_{\b{k}}\b{\delta}_{\b{k}}(t)\\
\b{\delta}_{\b{k}}(t)&=\left(\delta \sigma^x_{\b{k}}(t), \delta \sigma^y_{\b{k}}(t), \delta \sigma^z_{\b{k}}(t),\delta a(t), \delta a^{\dagger}(t)\right)^T
\end{align}
with the stability matrix 
\begin{align}
\mathcal{D}_{\b{k}}=\left(
\begin{array}{ccccc}
 -\frac{\gamma }{2} & \Delta -2 V (\braket{\sigma^z}+1) & -V \braket{\sigma^y} t_{\b{k}} & 0 & 0 \\
 2 V (\braket{\sigma^z}+1)-\Delta  & -\frac{\gamma }{2} & V \braket{\sigma^x} t_{\b{k}}-2 g (\braket{a}+\braket{a^\dagger}) & -2 g \braket{\sigma^z} & -2 g \braket{\sigma^z} \\
 0 & 2 g (\braket{a}+\braket{a^\dagger}) & -\gamma  & 2 g \braket{\sigma^y} & 2 g \braket{\sigma^y} \\
 -i g \delta(\b{k})  & 0 & 0 & -(\kappa +i \omega_0) & 0 \\
 +i g \delta(\b{k})  & 0 & 0 & 0 & -(\kappa -i \omega_0) \\
\end{array}
\right)
\label{EQ:StabilityMatrixAppendix}
\end{align}
here the momentum dependence is given by $t_{\b{k}}=\cos(k_x)+\cos(k_y)$.
The stability matrix has eigenvalues $\lambda$ that depend on the wave number $\b{k}$. The sign $(\pm)$ of the real part of the eigenvalues determine if perturbations with momentum $\b{k}$ decay (-) or grow (+) in time. If an eigenvalue acquires a positive real part, the uniform solution is unstable. The dynamics of the instability will be dominated by the wave vector $\b{k}$ for which ${\rm Re[\lambda]}$ is at its maximum. 
Inspecting the matrix in Eq.\,\eqref{EQ:StabilityMatrixAppendix}, one can see that for an infinite system size, it depends continuously on the momentum $\b{k}$ only through the Rydberg interaction which, on a mean-field level, favors ordering around $(k_x,k_y)=(\pi,\pi)$. The appearance of the delta function $\delta(\b{k})$ shows that fluctuations in the coherent photon field only couple to uniform perturbations. In particular, there is no competition with other $\b{k}$-dependent terms that could induce instabilities at finite momentum $\b{k}\neq(0,0)$.
In Fig.~\ref{Fig:Stabilityanalysis} we show where the homogeneous solution to Eqs.~(\ref{EQ:SigmaxApp}-\ref{EQ:adagApp}) (excluding the empty cavity, where $\braket{\sigma^x}=\braket{\sigma}^y=0$ and $\braket{\sigma^z}=-1$ and $\braket{a}$=0) in linear response is maximally unstable towards excitations at $(k_x,k_y)=(\pi,\pi)$. Within the even-odd sublattice Ansatz of Eqs.~(\ref{EQ:Sigmax1}-\ref{EQ:PhotonTerm}) we include the phase boundary of the stable $\rm{(AFM+SR)}$ solutions (solid line). Our results are fully consistent with each other. As it can be seen in Fig.\,\ref{Fig:Stabilityanalysis}, there is a region where the homogeneous solution is unstable towards excitation at $\b{k}=(\pi,\pi)^T$ but where the corresponding antiferromagnetic solution is not a stable steady-state. The entire phase-diagram is given in Fig.\,\ref{Fig:DRPhaseDiagramkappagamma}. 

%%%%%%%%%%%%%%%%%%%%%%%%%%%%%%%%%%%%%%%

\section{Hierarchy of energy scales and problematic Rydberg decays}
\label{app:blackbodydecays}

We now compare typical timescales associated to the Hamiltonian and Liouvillian dynamics given by Eqs.\,\eqref{eq:H_spin}-\eqref{EQ:lindbladphoton}, using two two recently performed experiment. One on a 
2d Ising Hamiltonian with an interaction between Rydberg-dressed ground states, (see Eq.\,\eqref{eq:H_spin}) carried 
out by Zeiher \textit{et al.\,} \cite{zeiher16} and an experiment by Baden \textit{et al.\,} \cite{baden14} with cavity-assisted Raman processes to realise the Dicke superradiance transition with ultracold atoms coupled to a high-finesse optical 
cavity, as described with the Hamiltonian given by Eq.\,\eqref{eq:H_spin-light}. 
The list of time and frequency scales is given in Table \ref{Tab:TimeScales} and in Table \ref{Tab:Frequencies}, respectively. 
\begin{table}
\begin{center}
\begin{tabular}{ c | c | c | c | c | c | c | c | c }
 & $\gamma_{\rm BB}\beta^2/2\pi$ & $\gamma_{r}\beta^2/2\pi$ & $V/2\pi$ & $\Delta^{\rm Z}/2\pi$ & $\kappa/2\pi$ & $\Delta^B/2\pi$ & $g_c/2\pi$ & $\omega_0/2\pi$ \\ \hline
kHz &0.003-0.020 & 0.06-0.45 &0.1-1.8 &27-64 & 100 & 50-100 & 50-150 & 100-300 
\end{tabular}
\caption{
Hierarchy of frequencies for all involved energy scales. The energy scales involving the spin-spin dynamics 
$(\gamma_{\rm BB},\gamma_r,V,\Delta^Z)$ are calculated from experiments by Zeiher \textit{et al.} \cite{zeiher16}. The energy scales ($\kappa,\Delta^B, g_c,\omega_0$) involving the spin-light and cavity dynamics are calculated from the experiments performed by Baden \textit{et al.} \cite{baden14}. Here, $\Delta^Z$ and $\Delta^B$ refer to the level splitting of the two-level atom and $\omega_0$ is the effective cavity detuning. $g_c$ refers to the critical atom-light coupling for the superradiance transition in the Singapore experiment. $\gamma_{\rm BB}$ and $\gamma_{r}$ refer to black-body radiation induced decay 
of the Rydberg-state \cite{Beterov09} and the decay time of the bare Rydberg state, respectively.}
\label{Tab:Frequencies}
\end{center}
\begin{center}
\begin{tabular}{ c | c | c | c | c | c | c | c | c }
 & $\tau_{\rm BB}/\beta^2$ & $\tau_{r}/\beta^2$ & $\tau_{V}$ & $\tau^{Z}_{\Delta}$ & $\tau_{\kappa}$ & $\tau^B_{\Delta}$ & $\tau_{g_c}$ & $\tau_{\omega_0}$ \\ \hline
$\mu$s &50880-361808 & 2200-15630 &552-10472 &15-36 & 10 & 6-20 & 10-20 & 3-10 
\end{tabular}
\caption{
Hierarchy of timescales for all involved processes calculated from table \ref{Tab:Frequencies}.
}
\label{Tab:TimeScales}
\end{center}
\end{table}
%
%As our model combines both experimentally used quantum optical implementation schemes we consider this a reasonable %starting point. 
It can be seen that the Rydberg-dressed interaction $V$ is relatively small compared to the other appearing energy scales. For an experimental realisation of a phase with an even/odd asymmetry it would thus be required to increase the strength of interaction. This can be achieved by reducing the laser detuning to the bare Rydberg level. However, this will lead to higher inherited loss rates for the admixed state. Additionally, it could be possible to prepare an initial many-body state such that it is close to a state with an even/odd symmetry breaking. A scheme to prepare such states in extended Rydberg ensembles is in Ref.~\onlinecite{Pohl2010}. As indicated in Table \ref{Tab:TimeScales}, radiative losses set the longest timescale of the system. However, 
blackbody radiation induced losses can limit the coherence time in Rydberg-dressing schemes 
\cite{zeiher16,Beterov09,Goldschmidt16} as a single decay event can lead to avalanche-like losses of atoms 
from the trapping lattice. However, it was also pointed out \cite{zeiher16}, that such impurity Rydberg atoms 
could be eliminated in future experiments by using a laser quench before atom-loss occurs. 

Specifically, $\gamma_{\rm BB}$ and $\gamma_{r}$ refer to black-body radiation induced decay 
of the Rydberg-state \cite{Beterov09} and the decay time of the bare Rydberg state, respectively. 
Both rates are modified 
by a dressing factor denoted by $\beta^2$ that determines the strength of admixing the Rydberg-level to the ground-state \cite{zeiher16}. In both tables, $\beta^2 \in (0.012-0.0017)$.
While spontaneous emission occurs predominantly to other ground or low-lying excited states, blackbody radiation transfers population from a virtual Rydberg excitation mostly to neighbouring high-lying states, $n\to n\pm 1$, where then a true Rydberg atom is created with a rate $\beta^2 \gamma_{\rm BB}$. By assuming stochastically triggered, instantaneous loss of all atoms in the state $\ket{\uparrow}$ good agreement with experimental data was obtained with a model to estimate the mean number $N(t)$ of remaining atoms after a dressing laser had been applied for a time $t$:
\begin{align}
N(t)\approx N(0)P(t), \quad P(t)\approx\exp\left(-\frac{N(0)}{4}\gamma_{\rm BB}\beta^2 t\right),
\end{align}
where $P(t)$ is the probability that no atom induced a blackbody-radiation induced loss process.
An experimental determination yielded a value of $\gamma_{\rm BB}/2\pi=1.6 kHz$ which was almost half the literature value for $31P_{1/2}$ states \cite{Beterov09}.

%The ensuing strong dipole-dipole interaction with other high-lying states can trigger an avalanche-like atom loss process of %all atoms that are in the Rydberg-dressed state \cite{Goldschmidt16}. Therefore, concerns have been raised that blackbody %radiation might put constrains on the time for coherent interactions between dressed Rydberg ground-states. At room %temperature, blackbody radiation is the dominating process to shorten radiative lifetimes for Rydberg states \cite{Beterov09}. %However in their conclusion, Zeiher \textit{et al.} mention that the avalanche atom-loss process might be prevented if an %appropriately adjusted Rydberg-state detuning is used, or a stroboscopic dressing scheme is employed instead, to laser %quench impurity Rydberg atoms.  

%%%%%%%%%%%%%%%%%%%%%%%%%%%%%%%%%%%%
%\newpage
%ieeetr
%\bibliographystyle{IEEEtran}
\bibliographystyle{apsrev4-1_custom}
\bibliography{RDicke-bibfile}
%\bibliographystyle{nature}
%\bibliography{dicke-bibfile}
%\bibliographystyle{nature}
%\bibliographystyle{aipnum4-1}

% alternative aipnum4-1

%\end{thebibliography}

\end{document}